\documentclass[preprints,article,accept,moreauthors,10pt,a4paper]{mdpi}

\usepackage{hyperref}
\newcommand\beq{\begin{equation}}
\newcommand\eeq{\end{equation}}
\newcommand\beqa{\begin{eqnarray}}
\newcommand\eeqa{\end{eqnarray}}

\newcommand{\al}{\alpha}

\firstpage{1} 
\pubvolume{1}
\issuenum{1}
\articlenumber{0}
\pubyear{2024}
\copyrightyear{2024}
\history{Received: date; Accepted: date; Published: date} 
\Title{Exact results for non-Newtonian transport properties in sheared granular suspensions: inelastic Maxwell models and BGK-type kinetic model}
\Author{Rub\'en G\'omez Gonz\'alez $^{1,\dagger}$\orcidA{}, Vicente Garz\'{o} $^{2,\dagger}$\orcidB{}}
\AuthorNames{Rub\'en G\'omez Gonz\'alez and Vicente Garz\'{o}}
\address{%
$^{1}$ \quad Departamento de F\'{\i}sica, Universidad de Extremadura, Avda. de Elvas s/n, E-06006 Badajoz, Spain; ruben@unex.es\\
$^{2}$ \quad Departamento de F\'{\i}sica and Instituto de Computaci\'on Cient\'{\i}fica Avanzada (ICCAEx), Universidad de Extremadura, Avda. de Elvas s/n, E-06006 Badajoz, Spain; vicenteg@unex.es}

\corres{Correspondence: vicenteg@unex.es; URL: https://fisteor.cms.unex.es/investigadores/vicente-garzo-puertos/}
\firstnote{These authors contributed equally to this work.}

\abstract{
The Boltzmann kinetic equation for dilute granular suspensions under simple (or uniform) shear flow (USF) is considered to determine the non-Newtonian transport properties of the system. In contrast to previous attempts based on a coarse-grained description, our suspension model accounts for the real collisions between grains and particles of the surrounding molecular gas. The latter is modeled as a bath (or thermostat) of elastic hard spheres at a given temperature. Two independent but complementary approaches are followed to reach exact expressions for the rheological properties. First, the Boltzmann equation for the so-called inelastic Maxwell models (IMM) is considered. The fact that the collision rate of IMM is independent of the relative velocity of the colliding spheres allows us to exactly compute the collisional moments of the Boltzmann operator without the knowledge of the distribution function. Thanks to this  property the transport properties of the sheared granular suspension can be \emph{exactly} determined. As a second approach, a Bhatnagar--Gross--Krook (BGK)-type kinetic model adapted to granular suspensions is solved to compute the velocity moments and the velocity distribution function of the system. The theoretical results (which are given in terms of the coefficient of restitution, the reduced shear rate, the reduced background temperature, and the diameter and mass ratios) show in general a good agreement with the approximate analytical results derived for inelastic hard spheres (IHS) by means of Grad's moment method and with computer simulations performed in the Brownian limiting case ($m/m_g\to \infty$, where $m_g$ and $m$ are the masses of the particles of the molecular and granular gases, respectively). In addition, as expected the IMM and BGK results show that the temperature and non-Newtonian viscosity exhibit and $S$ shape in a plane of stress-strain rate (discontinuous shear thickening, DST). The DST effect becomes more pronounced as the mass ratio $m/m_g$ increases.
}


\begin{document}
\section{Introduction}
\label{sec1}

A very usual way of assessing the effect of the surrounding fluid on the dynamics properties of solid particles is through an effective fluid-solid force \cite{K90,G94,FH17,J00}. In some models, this force is simply proportional to the velocity particle (Stokes linear drag law) \cite{TK95,SMTK96,WZLH09,H13,ChVG15,SA17,ASG19,SA20}. This type of force attempts to mimic the energy dissipated by grains due to their friction on the interstitial viscous gas. A more sophisticated model \cite{GTSH12,HTG17,HT19,GGG19,GGG19a,HTG20,THSG20,GKG20,THG23} incorporates also a Langevin-like stochastic term that accounts for the energy transferred to grains due to their ``collisions'' with particles of the background gas.       
However, although this coarse-grained approach has provided reliable results in the past, it would be desirable to consider a suspension model that takes into account the real collisions between grains and particles of the surrounding (molecular) gas. This sort of suspension model (which has been inspired in a previous work of Biben \emph{et al}. \cite{BMP02}) has been recently proposed \cite{GG22a}. In this model, granular particles are assumed to be sufficiently rarefied so that, they do not disturb the state of the molecular (background) gas. As a consequence, the interstitial gas may be treated as a \emph{thermostat} at the temperature $T_g$. Moreover, although the concentration (mole fraction) of grains is quite small, apart from the elastic collisions between solid and molecular gas particles one has to consider the \emph{inelastic} collisions between grains themselves. This model can be useful to analyze transport properties in particle-laden suspensions \cite{S20} where very dilute particles (like aerosols) are immersed in a fluid (like air).

The rheological properties of a granular suspension under simple (or uniform) shear flow (USF) has been recently determined \cite{GChG23}. In contrast to previous attempts \cite{TK95,SMTK96,WZLH09,H13,ChVG15,SA17,HTG17,HT19,ASG19,SA20,THSG20,THG23}, the results obtained in Ref.\ \cite{GChG23} have been derived from the collisional model proposed in Ref.\ \cite{GG22a}. On the other hand, a limitation of these results is that they have been \emph{approximately} obtained by employing Grad's moment method \cite{G49}, namely, a method based on the truncation of a series expansion of the velocity distribution function in (orthogonal) Sonine polynomials. The use of this approximate method is essentially motivated by the form of the collision rate for inelastic hard spheres (IHS) appearing inside the Boltzmann collision operator. The collision rate for IHS is proportional to the magnitude of the normal component of the relative velocity of the two spheres that are about to collide. This velocity dependence of the collision rate for IHS prevents the possibility of deriving exact expressions for the transport properties in the USF problem, even in the case of elastic collisions.

A possible way of overcoming the technical difficulty of the hard-sphere kernel is to consider the so-called inelastic Maxwell models (IMM). As for the conventional Maxwell molecules \cite{CC70,GS03}, the collision rate of IMM is independent of the relative velocity of the two colliding spheres \cite{BK03}. The use of IMM instead of IHS opens up the possibility of getting exact analytical results of the Boltzmann equation in some specific non-equilibrium situations, like the USF. In particular, the knowledge of the collisional moments of the Boltzmann equation for IMM \cite{GS07,SG23} enables a clear exploration of the impact of inelasticity on the non-Newtonian transport properties of the granular suspension without introducing uncontrolled approximations.

Another possible alternative for obtaining exact results is to consider a kinetic model that retains the relevant physical properties of the Boltzmann collision operator but turns out to be more tractable than the true kinetic equation. This route has been widely employed in the past in the case of molecular dilute gases \cite{GS03} where it has been shown that several exact solutions in far from equilibrium states agree very well with Monte Carlo simulations of the Boltzmann equation. Here, as in previous works \cite{GGG19a}, we will consider a kinetic model for granular suspensions \cite{VGS07} to complement the theoretical expressions obtained from the Boltzmann equation for IMM. Since this kinetic model is based on the well known Bhatnagar--Gross--Krook (BGK) model \cite{C88} for molecular gases, we will referred to it as a BGK-type kinetic model.

The objective of this paper is to determine the rheological properties of granular particles immersed in a bath of elastic hard spheres under USF. At a macroscopic level, the USF is characterized by constant number densities for solid and gas particles, a uniform temperature, and a (common) linear velocity profile $U_{g,x}=U_x=ay$, where $a$ is the constant shear rate. Here, $\mathbf{U}_g$ and $\mathbf{U}$ denote the mean flow velocities of the molecular and granular gases, respectively. Since we are interested here in the steady state where the system admits a non-Newtonian hydrodynamic description, an external thermostat force (proportional to the peculiar velocity) must be introduced to keep constant the temperature $T_g$ of the molecular gas.   

The use of IMM as well as a BGK-type kinetic model allows us to exactly compute the rheological properties of the granular suspension. These properties are expressed as nonlinear functions of the (reduced) shear rate $a^*=a/\gamma$ (where $\gamma$ is a drift coefficient characterizing the friction of solid particles on the viscous gas), the coefficient of restitution, the (reduced) background temperature $T_g^*$, and the diameter $\sigma/\sigma_g$ and mass $m/m_g$ ratios. Here, $\sigma_g$ and $m_g$ are the diameter and mass of the particles of the molecular gas, respectively, while $\sigma$ and $m$ are the diameter and mass of the solid particles, respectively. As occurs for IHS \cite{GChG23}, our results show that the kinetic granular temperature and the non-Newtonian viscosity exhibit a discontinuous shear thickening (DST) effect for sufficiently large values of the mass ratio $m/m_g$. In fact, in the Brownian limiting case ($m/m_g \to \infty$) the expressions of the rheological properties derived here reduce to those previously obtained \cite{GGG19a} from a coarse-grained description based on the Fokker--Planck operator. This agreement justifies the use of this latter approach to analyze the DST effect in dilute granular suspensions \cite{PZ23}.         

Apart from the transport properties (which are related with the second-degree velocity moments), the explicit forms of the higher degree velocity moments as well as the velocity distribution function of the granular gas have been also obtained from the BGK model. This is one of the main advantages of using a kinetic model instead of the true Boltzmann equation. Our results show in particular that the fourth-degree moments of the distribution function also exhibit a DST effect. With respect to the velocity distribution function, as expected we find that its distortion from equilibrium is more significant as both the mass $m/m_g$ and diameter $\sigma/\sigma_g$ ratios depart from 1. In addition, a comparison between the BGK results and numerical solutions of the Boltzmann equation from the direct simulation Monte Carlo (DSMC) method \cite{B94} for IHS shows a generally good qualitative agreement between both approaches. 
%However, some quantitative discrepancies arise, especially in the region of thermal velocities} ($|\mathbf{c}|\sim 1$, where $\mathbf{c}=\sqrt{m/2T_g}\mathbf{V}$). 

The plan of the paper is as follows. The Boltzmann kinetic equation for a granular gas immersed in a bath of elastic hard spheres under USF is presented in Section \ref{sec2}. The balance equations for the temperatures of the molecular and granular gases are also displayed. Section \ref{sec3} deals with the calculations carried out for IMM of the rheological properties of the granular suspension. While a shear thinning effect is always found for the nonlinear shear viscosity of the molecular gas, the corresponding shear viscosity of the granular gas exhibits a DST effect for sufficiently large values of the mass ratio $m/m_g$. The results derived from the BGK-type kinetic model are provided in Section \ref{sec4} while a comparison between the theoretical results obtained for IHS, IMM and BGK model is displayed in Section \ref{sec5} for several systems. Our results highlight a good agreement for the rheology between the three different approaches. Moreover, theoretical results obtained from IMM and BGK model are also compared against computer simulations in the Brownian limit ($m/m_g \to \infty$) showing a good agreement. The paper is closed in Section \ref{sec6} with some concluding remarks.

\section{Boltzmann kinetic equation for sheared granular suspensions}
\label{sec2}

We consider a set of solid particles (granular gas) of mass $m$ and diameter $\sigma$ which are immersed in a solvent (molecular gas) constituted by particles of mass $m_g$ and diameter $\sigma_g$. As usual, the granular gas is modeled as a gas of hard disks ($d=2$) or spheres ($d=3$) with inelastic collisions. In the simplest model, the inelasticity of collisions is characterized by a constant (positive) coefficient of normal restitution $\al\leq 1$, where $\al=1$ refers to elastic collisions. On the other hand, collisions between solid particles and particles of the molecular gas are elastic. We also assume that the number density of grains is much smaller than that of solvent so that, the state of the latter is not perturbed by the presence of the former. In these conditions, we can treat the molecular gas as a bath or \emph{thermostat} at the temperature $T_g$ (once the parameters of the system, specifically the shear rate, have been set). Moreover, although the granular gas is sufficiently rarefied, we take into account the collisions among grains in its corresponding Boltzmann kinetic equation.  

We assume that the system (granular particles plus solvent) is subjected to USF. As said in the Introduction section, this state is characterized by constant densities $n_g$ and $n$, uniform temperatures $T_g$ and $T$, and 
by a (common) linear profile of the $x$ component of the flow velocities along the $y$ axis: 
\beq
\label{2.1}
n_g\equiv \text{const}, \quad n\equiv \text{const},
\eeq
\beq
\label{2.2}
\nabla T_g=\nabla T=0,
\eeq
\beq
\label{2.3}
U_{g,i}=U_i=a_{ij}r_j, \quad a_{ij}=a\delta_{ix}\delta_{jy},
\eeq
$a$ being the \emph{constant} shear rate. Here, $n_g$, $\mathbf{U}_g$, and $T_g$ are the number density, the mean flow velocity, and the temperature, respectively, of the molecular gas. In terms of its one-particle velocity distribution function $f_g(\mathbf{r},\mathbf{v};t)$, these hydrodynamic fields are defined as  
\beq
\label{2.4}
\left\{n_g, n_g \mathbf{U}_g, d n_g T_g \right\}=\int d\mathbf{v}\left\{1, \mathbf{v}, m_g V^2\right\}f_g(\mathbf{v}),
\eeq
where $\mathbf{V}=\mathbf{v}-\mathbf{U}$ is the peculiar velocity. Note that in Eq.\ \eqref{2.4} the Boltzmann constant $k_\text{B}=1$. We will take this value throughout the paper for the sake of simplicity. In addition, in Eqs.\ \eqref{2.1}--\eqref{2.3}, $n$, $\mathbf{U}$, and $T$ denote the number density, the mean flow velocity, and the (granular) temperature, respectively, of the granular gas. They are defined as    
 \beq
\label{2.5}
\left\{n, n \mathbf{U}, d n T \right\}=\int d\mathbf{v}\left\{1, \mathbf{v}, m V^2\right\}f(\mathbf{v}).
\eeq 

Since the only spatial gradient present in the USF problem is the shear rate, the pressure tensor 
\beq
\label{2.6}
\mathsf{P}_g=m_g \int d\mathbf{v}\; \mathbf{V} \mathbf{V} f_g(\mathbf{v})
\eeq
of the molecular gas, and the pressure tensor
\beq
\label{2.6.1}
\mathsf{P}=m \int d\mathbf{v}\; \mathbf{V} \mathbf{V} f(\mathbf{v})
\eeq 
of the granular gas are the relevant fluxes in the problem. They provide information on the transport of momentum across the system. Our main target is to determine $\mathsf{P}_g$ and $\mathsf{P}$ for arbitrary shear rates. 

One of the main advantages of the USF at a microscopic level is that it becomes a spatially homogeneous state when the velocities of the particles are referred to a Lagrangian frame moving with the linear velocity $U_i=a_{ij}r_j$. In this new frame and in the steady state, the distribution functions of the molecular and granular gases adopt the form
\beq
\label{2.7}
f_g(\mathbf{r},\mathbf{v})=f_g(\mathbf{V}), \quad f(\mathbf{r},\mathbf{v})=f(\mathbf{V}).
\eeq
In addition, as the state of the solvent is not perturbed by the solid particles, the temperature $T_g$ in the USF state increases in time due to the viscous heating term $-aP_{xy}>0$. Thus, as usual in nonequilibrium molecular dynamics simulations \cite{EM90}, an external nonconservative force (thermostat) is introduced in the molecular gas to achieve a stationary state. Among the different possibilities, for simplicity, a force proportional to the particle velocity (Gaussian thermostat) of the form $\mathbf{F}_g=-m_g \xi \mathbf{V}$ is considered in this paper. The parameter $\xi$ is chosen to be a function of the shear rate by the condition that $T_g$ reaches a constant value in the long time limit. Analogously, the granular gas is also subjected to this kind of Gaussian thermostat (i.e., $\mathbf{F}=-m \xi \mathbf{V}$), where $\xi$ is the same quantity for the solvent and the solid particles. 

Under the above conditions, in the low-density regime, the distribution function $f_g(\mathbf{V})$ of the molecular gas obeys the nonlinear (closed) Boltzmann equation 
\beq
\label{2.8}
-a V_y \frac{\partial f_g}{\partial V_x}-\xi \frac{\partial}{\partial \mathbf{V}}\cdot 
 \mathbf{V} f_g=J_g[\mathbf{V}|f_g,f_g],
\eeq
while the distribution function $f(\mathbf{V})$ of the granular gas obeys the kinetic equation
\beq
\label{2.9}
-a V_y \frac{\partial f}{\partial V_x}-\xi \frac{\partial}{\partial \mathbf{V}}\cdot \mathbf{V} f=J[\mathbf{V}|f,f]+J_\text{BL}[\mathbf{V}|f,f_g].
\eeq
Here, $J_g[f_g,f_g]$ and $J[f,f]$ are the nonlinear Boltzmann collision operators for the molecular and granular gases, respectively, and $J_\text{BL}[f,f_g]$ is the linear Boltzmann--Lorentz collision operator \cite{BP04,G19}.    
The balance equations for $T_g$ and $T$ can easily obtained by multiplying both sides of Eqs.\ \eqref{2.8} and \eqref{2.9} by $m_g V^2$ and $m V^2$, respectively, and integrating over velocity. The results are
\beq
\label{2.10}
-aP_{g,xy}=d\xi p_g,
\eeq
\beq
\label{2.11}
-aP_{xy}-\frac{d}{2}p\zeta_g=d\xi p+\frac{d}{2}p\zeta,
\eeq
where $p_g=n_g T_g$ and $p=n T$ are the hydrostatic pressures of the molecular and granular gases, respectively, and the partial production rates $\zeta$ and $\zeta_g$ are defined, respectively, as
\beq
\label{2.12}
\zeta=-\frac{m}{d n T}\int d\mathbf{v}\; V^2\; J[\mathbf{v}|f,f], \quad
\zeta_g=-\frac{m}{d n T}\int d\mathbf{v}\; V^2\; J_\text{BL}[\mathbf{v}|f,f_g].
\eeq
The cooling rate $\zeta$ gives the rate of kinetic energy loss due to inelastic collisions between particles of the granular gas. It vanishes for elastic collisions $(\al=1)$. The term $\zeta_g$ gives the transfer of kinetic energy between the particles of the granular gas and the solvent. This term vanishes when the granular and molecular gases are at the same temperature ($T_g=T$). Equation \eqref{2.10} implies that in the steady state the viscous heating term ($-aP_{g,xy}>0$) is exactly balanced by the heat extracted in the gas by the external thermostat. On the other hand, since $\zeta_g$ can be positive or negative, Eq.\ \eqref{2.11} implies that in the steady state the term $-aP_{xy}-(d/2)p \zeta_g$ is exactly compensated for the cooling terms arising from collisional dissipation ($\zeta p$) and the thermostat term ($\xi p$). 
%Thus, at a given values of the diameter $\sigma_g/\sigma$ and mass $m_g/m$ ratios and the bakground temperature $T_g$, the temperature ratio $\chi=T/T_g$ is a function of both the coefficient of restitution $\al$ and the (reduced) shear rate $a^*=a/\gamma$, where $\gamma$ is an effective collision frequency to be defined later. 

The USF state is in general a non-Newtonian state characterized by shear-rate dependent transport coefficients. In particular, one can define the non-Newtonian shear viscosity of the molecular gas as
\beq
\label{2.13}
\eta_g=-\frac{P_{g,xy}}{a}. 
\eeq
Analogously, the non-Newtonian shear viscosity of the granular gas is given by 
\beq
\label{2.13.1}
\eta=-\frac{P_{xy}}{a}.
\eeq
In addition, beyond the Navier--Stokes domain, normal stress differences are expected in the USF. This means that $P_{g,xx}\neq P_{g,yy} \neq P_{g,zz}$ and $P_{xx}\neq P_{yy} \neq P_{zz}$. 

It is quite evident that the evaluation of the rheological properties of the molecular and granular gases requires the knowledge of the pressure tensors $\mathsf{P}_g$ and $\mathsf{P}$. The non-zero elements of these tensors can be obtained by multiplying by $m_g \mathbf{V}\mathbf{V}$ and $m \mathbf{V}\mathbf{V}$ both sides of  Eqs.\ \eqref{2.8} and \eqref{2.9}, respectively, and integrating over $\mathbf{V}$. However, to achieve explicit forms for  $\mathsf{P}_g$ and $\mathsf{P}$, one has to compute the collisional moments
\begin{equation}
\label{2.14}
\mathsf{A}_g=m_g\int d{\bf V} {\bf V}{\bf V} J_{g}[f_g,f_g],
\end{equation}
\begin{equation}
\label{2.15}
\mathsf{B}=m\int d{\bf V} {\bf V}{\bf V} J[f,f], \quad 
\mathsf{C}=m\int d{\bf V} {\bf V}{\bf V} J_\text{BL}[f,f_g].
\end{equation}

In the case of IHS, the collisional moments $\mathsf{A}_g$, $\mathsf{B}$, and $\mathsf{C}$ cannot be exactly computed. As said in the Introduction section, a good estimate of them for IHS has been made in Ref.\ \cite{GChG23} by means of Grad's moment method \cite{G49}. This method is based on the expansion of the distributions $f_g(\mathbf{V})$ and $f(\mathbf{V})$ in a complete set of orthogonal polynomials; the coefficients being the corresponding velocity moments of those distributions. The above expansion generates an infinite hierarchy of moment equations that must be truncated at a given order. This truncation allows one to arrive to a closed set of coupled equations for the velocity moments that can be recursively solved. Thus, since the results derived in Ref.\ \cite{GChG23} for the rheological properties of molecular and granular gases are approximated, it is convenient to revisit the problem and determine the \emph{exact} expressions for the non-Newtonian transport properties of the granular suspension. To achieve such exact forms, two independent approaches will be considered in this paper: (i) the Boltzmann kinetic equation for IMM and (ii) a BGK-type kinetic model for IHS. This task will be carried out in the next two Sections.         

\section{Rheology from inelastic Maxwell models}
\label{sec3}

In this Section, we will consider IMM, namely, a collisional model where the collision rate of the two colliding spheres are independent of their relative velocity. In this case, the Boltzmann collision operator $J_g[f_g,f_g]$ \footnote{This is a simple version of the Boltzmann collision operator for Maxwell molecules} of the molecular gas can be written as \cite{C88,GS03}
\beq
\label{3.1}
J_g[\mathbf{v}_1|f_g,f_g]=\frac{\nu_g^\text{M}}{n_g S_d}\int d\mathbf{v}_2 \int d\widehat{\boldsymbol {\sigma}}\,
\left[f_{g}({\bf V}_{1}'')f_{g}({\bf V}_{2}'')-f_{g}({\bf V}_1)f_{g}({\bf V}_{2})\right],
\eeq
where $S_d=2\pi^{d/2}/\Gamma(\frac{d}{2})$ is the total solid angle in $d$ dimensions and $\nu_g^\text{M}$ is an independent-velocity collision
frequency. In Eq.\ \eqref{3.1}, the primes on the velocities denote the initial values $\{{\bf V}_{1}^{\prime\prime},{\bf V}_{2}^{\prime\prime}\}$ that lead to $\{{\bf V}_{1},{\bf V}_{2}\}$ following a binary collision:
\begin{equation}
\label{3.2}
\mathbf{V}_{1}^{''}=\mathbf{V}_{1}- (\widehat{\boldsymbol {\sigma}}\cdot 
{\bf g}_{12})\widehat{\boldsymbol {\sigma}}, \quad 
\mathbf{V}_{2}^{''}=\mathbf{V}_{2}+(\widehat{\boldsymbol {\sigma}}\cdot 
{\bf g}_{12})\widehat{\boldsymbol {\sigma}}.
\end{equation}
The effective collision frequency $\nu_g^\text{M}$ can be seen as a free parameter of the model to be chosen to get agreement with the properties of interest of the original Boltzmann equation for IHS. For instance, to correctly capture the velocity dependence of the original IHS collision rate, we can assume that the IMM collision rate is proportional to $\sqrt{T}_g$.  

In the context of IMM, the inelastic Boltzmann collision operator $J[f,f]$ is \cite{BK03,G19}
\begin{equation}
\label{3.3}
J[\mathbf{V}|f,f]=
\frac{\nu^\text{M}}{n S_d}\int d{\bf V}_{2}\int d\widehat{\boldsymbol {\sigma}}\,
\left[\al^{-1}f({\bf V}_{1}'')f({\bf V}_{2}'')-f({\bf V}_1)f({\bf V}_{2})\right],
\end{equation}
while the Boltzmann--Lorentz collision operator $J_\text{BL}[f,f_g]$ is defined as \cite{C88,GS03}  
\beq
\label{3.4}
J_\text{BL}[\mathbf{V}|f,f_g]=
\frac{\nu_0^\text{M}}{n S_d}\int d{\bf V}_{2}\int d\widehat{\boldsymbol {\sigma}}\,
\left[f({\bf V}_{1}'')f_g({\bf V}_{2}'')-f({\bf V}_1)f_g({\bf V}_{2})\right].
\eeq
The relationship between $(\mathbf{V}_1^{\prime\prime}, \mathbf{V}_2^{\prime\prime})$ and $(\mathbf{V}_1, \mathbf{V}_2)$ in Eq.\ \eqref{3.3} is
\beq
\label{3.5}
\mathbf{V}_{1}''=\mathbf{V}_{1}-\frac{1+\al}{2\al}(\widehat{{\boldsymbol {\sigma}}} \cdot {\mathbf g}_{12})\widehat{\boldsymbol {\sigma}}, \quad 
\mathbf{V}_{2}''=\mathbf{V}_{2}+\frac{1+\al}{2\al}(\widehat{{\boldsymbol {\sigma}}} \cdot {\mathbf g}_{12})\widehat{\boldsymbol {\sigma}},
\eeq
while in Eq.\ \eqref{3.4} is
\beq
\label{3.6}
\mathbf{V}_{1}''=\mathbf{V}_{1}-2 \mu_g(\widehat{{\boldsymbol {\sigma }}} \cdot {\mathbf g}_{12})\widehat{\boldsymbol {\sigma }}, \quad \mathbf{V}_{2}''=\mathbf{V}_{2}+2 \mu(\widehat{{\boldsymbol {\sigma }}} \cdot {\mathbf g}_{12})\widehat{\boldsymbol {\sigma}}
\eeq
where 
\beq
\label{3.7}
\mu_g=\frac{m_g}{m+m_g}, \quad  \mu=\frac{m}{m+m_g}.
\eeq
In addition, as in the case of the collision frequency $\nu_g^\text{M}$, the collision frequencies $\nu^\text{M}$ and $\nu_0^\text{M}$ for granular-granular and granular-molecular collisions, respectively, can be chosen to optimize the agreement with the results derived from IHS. We will chose them later. 

As mentioned in previous works on IMM \cite{SG07,GS07,GGG19}, the main advantage of computing the collisional moments of the Boltzmann operator for Maxwell models (both elastic and inelastic models) is that they can be exactly provided in terms of the velocity moments of the distribution functions without the explicit knowledge of the latter. This property has been exploited to compute the second, third, and fourth-degree collisional moments of IMM for monocomponent \cite{GS07} and multicomponent \cite{G03bis,SG23} granular gases. The exact knowledge of the second-degree collisional moments allow us to get exact expressions for the rheological properties of the molecular and granular gases. Let us evaluate separately the rheology of both gases.          

\subsection{Rheological properties of the molecular gas}

The pressure tensor of the molecular gas is defined by Eq.\ \eqref{2.6}. To obtain the non-zero elements of this tensor, one multiplies both sides of the Boltzmann equation \eqref{2.8} by $m_g V_i V_j$ and integrates over velocity. The result is  
\begin{equation}
\label{3.8}
a_{ik}P_{g,kj}+a_{j k}{P}_{g,ki}+2\xi P_{g,ij}=-\frac{2\nu_g^\text{M}}{d+2} \left({P}_{g,ij}-p_g \delta_{ij}\right),
\end{equation}
where $p_g=n_gT_g$ is the hydrostatic pressure of the molecular gas. Upon obtaining Eq.\ \eqref{3.8}, use has been made of the result \cite{GS07}
\begin{equation}
\label{3.9}
{A}_{g,ij}=m_g\int d{\bf V} V_i V_j J_{g}[f_g,f_g]=-\frac{2\nu_g^\text{M}}{d+2} \left({P}_{g,ij}-p_g \delta_{ij}\right).
\end{equation}

The (reduced) elements of the pressure tensor $\mathsf{P}_g^*=\mathsf{P}_g/(n_g T_g)$ can be easily obtained from Eq.\ \eqref{3.8}. They are given by
\begin{equation}
\label{3.10}
P_{g,xx}^*=\frac{1}{1+2\widetilde{\xi}}\left[1+\frac{2\widetilde{a}^2}{(1+2\widetilde{\xi})^2}\right], \quad
P_{g,yy}^*=P_{g,zz}^*=\frac{1}{1+2\widetilde{\xi}}, \quad P_{g,xy}^*=-\frac{\widetilde{a}}{(1+2\widetilde{\xi})^2}.
\end{equation}
Here, we have introduced the quantities
\beq
\label{3.11}
\widetilde{a}=\frac{a}{\widetilde{\nu}_g^\text{M}}, \quad \widetilde{\xi}=\frac{\xi}{\widetilde{\nu}_g^\text{M}}, \quad \widetilde{\nu}_g^\text{M}=\frac{2}{d+2}\nu_g^\text{M}.
\eeq
The constraint $P_{g,xx}^*+(d-1)P_{g,yy}^*=d$ leads to a cubic equation relating $\widetilde{\xi}$ and $\widetilde{a}$:
\begin{equation}
\label{3.12}
\widetilde{a}^2=d\widetilde{\xi}(1+2\widetilde{\xi})^2.
\end{equation}
The real root of Eq.\ \eqref{3.12} gives the shear-rate dependence of $\widetilde{\xi}(\widetilde{a})$. It is given by \cite{IT56,GS03}
\beq
\label{3.13}
\widetilde{\xi}(\widetilde{a})=\frac{2}{3}\sinh^2\left[\frac{1}{6}\cosh^{-1}\left(1+\frac{27}{d}\widetilde{a}^2\right)\right].
\eeq

Comparison between the results derived here for $P_{g,ij}^*$ with those recently \cite{GChG23}  obtained for IHS by means of Grad's moment method \cite{G49} shows that both results are identical if the effective collision frequency $\nu_g^\text{M}$ is given by   
\beq
\label{3.14}
\nu_g^\text{M}=\frac{4\pi^{(d-1)/2}}{\Gamma\left(\frac{d}{2}\right)}n_g \sigma_g^{d-1}\sqrt{\frac{T_g}{m_g}}.
\eeq
Henceforth, we will take the choice \eqref{3.14} for $\nu_g^\text{M}$.

From Eqs.\ \eqref{3.10}, one can identify the (dimensionless) non-Newtonian shear viscosity $\eta_g^*=\nu_g^\text{M} \eta_g/p_g=-P_{g,xy}^*/\widetilde{a}$ and the (dimensionless) normal stress difference $\Psi_g^*=P_{g,xx}^*-P_{g,yy}^*$ as 
\beq
\label{3.15}
\eta_g^*=\frac{1}{(1+2\widetilde{\xi})^2}, \quad \Psi_g^*=\frac{2\widetilde{a}^2}{(1+2\widetilde{\xi})^3}.
\eeq
Note that the results derived here for Maxwell molecules yield $P_{g,yy}=P_{g,zz}$. This result contrasts with the one obtained for hard spheres by numerically solving the Boltzmann equation by means of the DSMC method \cite{B94} where it has been shown that $P_{g,yy}\neq P_{g,zz}$. However, the difference $P_{g,yy}- P_{g,zz}$ found in the simulations is in general quite small \cite{ChVG15}.  

It is also important to remark that in the case of Maxwell molecules there is an exact equivalence between the description with and without the drag force $-m \xi \mathbf{V}$. Nevertheless, for non-Maxwell molecules this type of force does not play a neutral role in the
transport properties of the system \cite{DSBR86}. 
\begin{figure}[ht]
\begin{center}
\includegraphics[width=0.5\columnwidth]{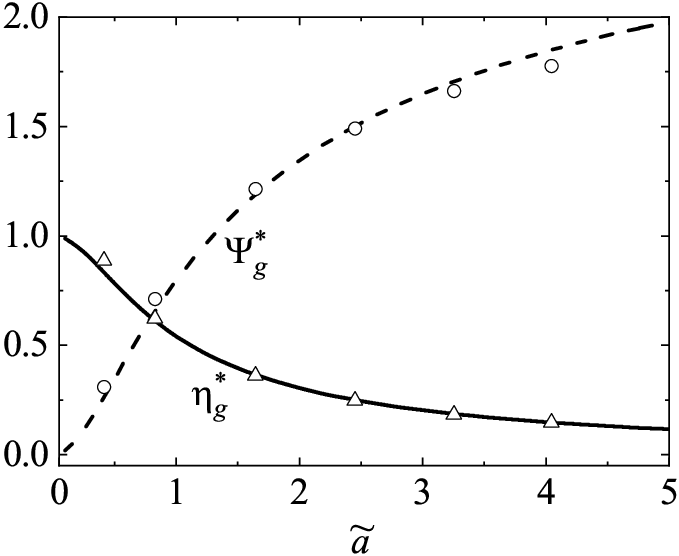}
\end{center}
\caption{Plot of the nonlinear shear viscosity $\eta_g^*$ and the normal stress difference $\Psi_g^*$ for hard spheres ($d=3$) as functions of the (reduced) shear rate $\widetilde{a}$. Symbols refer to the DSMC results for hard spheres. 
\label{fig0}}
\end{figure}
The shear-rate dependence of $\eta_g^*$ and $\Psi_g^*$ is plotted in Fig.\ \ref{fig0} for a three-dimensional system ($d=3$). As expected, the nonlinear viscosity $\eta_g^*$ decreases with increasing the (reduced) shear rate $\widetilde{a}$ (shear thinning effect). The opposite effect is observed for the normal stress difference function $\Psi_g^*$ since it increases with the shear rate. Figure \ref{fig0} also highlights the excellent agreement found between the theoretical results for IMM with those obtained by numerically solving the Boltzmann equation for hard spheres from the DSMC method \cite{B94}.

\subsection{Rheological properties of the granular gas}

As in the case of the molecular gas, the rheology of the granular gas can be also determined by multiplying both sides of Eq.\ \eqref{2.9} by $m V_i V_j$ and integrating over $\mathbf{V}$. After some algebra, one achieves the result
\beqa
\label{3.16}
& & a_{ik}P_{kj}+a_{j k}{P}_{ki}+2\xi P_{ij}=-\nu^\text{M}  \left[\nu_\eta^* P_{ij}+p\left(\zeta^*-\nu_\eta^*\right) \delta_{ij}\right]-\frac{4}{d} \nu_0^\text{M} \mu_g \nonumber\\
& & \times \Big[\Big(1-\frac{2}{d+2}\mu_g\Big)P_{ij}-\frac{2}{d+2}\mu \frac{n}{n_g}P_{g,ij}-\frac{d}{d+2}\mu_g n T_g\Big(\chi+\frac{m}{m_g}\Big)\delta_{ij}\Big],
\eeqa
where use has been made of the results \cite{GS07,G03bis,SG23}
\beq
\label{3.17}
B_{ij}=m\int d{\bf V} V_i V_j J[f,f]=-\nu^\text{M}  \left[\nu_\eta^* P_{ij}+p\left(\zeta^*-\nu_\eta^*\right) \delta_{ij}\right],
\eeq
\beqa
\label{3.18}
C_{ij}&=&m\int d{\bf V} V_i V_j J_\text{BL}[f,f_g]=-\frac{4}{d} \nu_0^\text{M} \mu_g\Big[\Big(1-\frac{2}{d+2}\mu_g\Big)P_{ij}-\frac{2}{d+2}\mu \frac{n}{n_g}P_{g,ij}\nonumber\\
& &-\frac{d}{d+2}\mu_g n T_g\Big(\chi+\frac{m}{m_g}\Big)\delta_{ij}\Big].
\eeqa
In Eq.\ \eqref{3.16}, 
\beq
\label{3.18.1}
\zeta^*=\frac{\zeta}{\nu^\text{M}}=\frac{1-\al^2}{2d}
\eeq
is the (reduced) cooling rate for the granular gas, $\chi=T/T_g$ is the temperature ratio, and 
\beq
\label{3.19}
\nu_\eta^*=\frac{\left(d+1-\al\right)(1+\al)}{d(d+2)}.
\eeq
The partial cooling rate $\zeta_g$ can be exactly obtained from Eq.\ \eqref{3.17} as 
\beq
\label{3.20}
\zeta_g=\frac{4\nu_0^\text{M}}{d}\mu_g\left[1-\mu_g\left(1+\theta\right)\right],
\eeq
where 
\beq
\label{3.20.1}
\theta=\frac{m T_g}{m_g T}
\eeq
is the ratio of the mean square velocities of granular and molecular gas particles. The forms \eqref{3.18.1} and \eqref{3.20} can be employed to fix the values of the free parameters $\nu^\text{M}$ and $\nu_0^\text{M}$. They are chosen under the criterion that $\zeta$ and $\zeta_g$ of IMM are the same as that of IHS of diameters $\sigma$ and $\sigma_0$. In this latter case, the above cooling rates are estimated by using Grad's approximation \cite{GChG23}. In this approximation, 
\beq
\label{3.21}
\zeta^{\text{IHS}}=\frac{2\pi^{(d-1)/2}}{d\Gamma\left(\frac{d}{2}\right)}n \sigma^{d-1}\sqrt{\frac{T}{m}}(1-\al^2), 
\eeq
\beq
\label{3.21.1}
\zeta_g^{\text{IHS}}=\frac{8\pi^{(d-1)/2}}{d\Gamma\left(\frac{d}{2}\right)}n_g \overline{\sigma}^{d-1}
\mu_g \left(\frac{1+\theta}{\theta}\right)^{1/2}\sqrt{\frac{2T_g}{m_g}}\left[1-\mu_g\left(1+\theta\right)\right],
\eeq
where $\overline{\sigma}=(\sigma+\sigma_g)/2$. Equations \eqref{3.20}, \eqref{3.21}, and \eqref{3.21.1} yield the identities
\beq
\label{3.22}
\nu^\text{M}=\frac{4\pi^{(d-1)/2}}{\Gamma\left(\frac{d}{2}\right)}n \sigma^{d-1}\sqrt{\frac{T}{m}}, \quad 
\nu_0^\text{M}=\frac{2\pi^{(d-1)/2}}{\Gamma\left(\frac{d}{2}\right)}n_g \overline{\sigma}^{d-1}\left(\frac{1+\theta}{\theta}\right)^{1/2}\sqrt{\frac{2T_g}{m_g}}.
\eeq

To compare with the rheological properties of IHS \cite{GChG23}, it is convenient at this level of the description to identify the friction (or drift) coefficient $\gamma$ appearing in the Brownian limiting case ($m/m_g\to \infty$) when the molecular gas is at equilibrium. In fact, this limiting case is the situation considered when one employs 
a coarse-grained approach \cite{K90,G94,J00,FH17} to assess the impact of the interstitial gas on the dynamics properties of grains. In this limiting case, the expression \eqref{3.18} of the collisional moment $C_{ij}$ reduces to   
\beq
\label{3.23}
C_{ij}^{\text{Br}}=-\frac{8\pi^{(d-1)/2}}{d\Gamma\left(\frac{d}{2}\right)}n T_g n_g \overline{\sigma}^{d-1}\left(\frac{m_g}{m}\right)^{1/2}\sqrt{\frac{2T_g}{m}}\left(P_{k\ell}^*-\delta_{k\ell}\right),
\eeq
where $P_{ij}^*=P_{ij}/(n T_g)$ and we have taken into account that in the Brownian limit $\mu_g\to m_g/m$ and  $(1+\theta)/\theta\to 1$ in the expression  \eqref{3.22} of $\nu_0^\text{M}$. The form of $C_{ij}$ derived in Ref.\ \cite{GGG19a} by replacing the Boltzmann--Lorentz collisional operator \eqref{3.4} by the Fokker--Planck operator 
\beq
\label{3.24}
\gamma \frac{\partial}{\partial \mathbf{V}}\cdot \mathbf{V}f+\gamma \frac{T_g}{m} \frac{\partial^2 f}{\partial V^2}
\eeq
is
 \beq
\label{3.24.1}
C_{ij}=-2 \gamma n T_g \left(P_{k\ell}^*-\delta_{k\ell}\right).
\eeq
Comparison between Eqs.\ \eqref{3.23} and \eqref{3.24.1} allows us to identify $\gamma$ for IMM as
\beq
\label{3.25}
\gamma=\frac{4\pi^{(d-1)/2}}{d\Gamma\left(\frac{d}{2}\right)}n_g \overline{\sigma}^{d-1}
\left(\frac{m_g}{m}\right)^{1/2}\left(\frac{2T_g}{m}\right)^{1/2}.
\eeq
The expression \eqref{3.25} for the friction coefficient $\gamma$ for IMM is the same as the one obtained for IHS \cite{GG22a,GChG23}.

We are now in conditions to obtain the nonzero elements of the (reduced) pressure tensor $P_{ij}^*$. From Eqs.\ \eqref{3.16} and  \eqref{3.25}, one gets the equation
\beqa
\label{3.26}
a_{ik}^*P_{kj}^*+a_{jk}^*{P}_{ki}^*+2\xi^* P_{ij}^*&=&-\nu^{*\text{M}} \left[\nu_\eta^*P_{ij}^*+\chi (\zeta^*-\nu_\eta^*)\delta_{ij}\right]
-2 \mu \left(\frac{1+\theta}{\theta}\right)^{1/2}\nonumber\\
& & \times \left(X \delta_{ij}+ Y P_{ij}^*+Z P_{g,ij}^*\right),
\eeqa
where $a_{ij}^*=a_{ij}/\gamma$, $\xi^*=\xi/\gamma$,
\beq
\label{3.27}
\nu^{*\text{M}}=\frac{\nu^\text{M}}{\gamma}=\frac{2^{d+1}d}{\sqrt{\pi}}\phi \sqrt{\chi T_g^*},
\eeq
and 
\beq
\label{3.28}
X=-\frac{d}{d+2}\mu \left(\frac{1+\theta}{\theta}\right), \quad Y=1-\frac{2\mu_g}{d+2}, \quad Z=-\frac{2\mu}{d+2}.
\eeq
Here, 
\beq
\label{3.29}
\phi=\frac{\pi^{d/2}}{2^{d-1}d \Gamma\left(\frac{d}{2}\right)}
n\sigma^d
\eeq
is the solid volume fraction of the granular gas, $T_g^*=T_g/m\sigma^2\gamma^2$, and upon deriving Eq.\ \eqref{3.27} use has been made of the identity
\beq
\label{3.30}
\frac{n\sigma^{d-1}}{n_g\overline{\sigma}^{d-1}}=\frac{2^{d+\frac{3}{2}}}{\sqrt{\pi}}\left(\frac{m_g}{m}\right)^{1/2}\phi \sqrt{T_g^*}.
\eeq

As occurs for the rheology of the molecular gas, Eq.\ \eqref{3.26} shows that the diagonal elements of the pressure tensor $P_{ij}^*$ orthogonal to the shear plane $xy$ are equal to $P_{yy}^*$ (i.e., $P_{yy}^*=P_{zz}^*=\cdots=P_{dd}^*$). This implies that the $xx$ element is given by $P_{xx}^*=d \chi-(d-1)P_{yy}^*$. The $yy$ and $xy$ elements of the (reduced) pressure tensor can be written as
\beq
\label{3.31}
P_{yy}^*=\frac{\Omega_{yy}}{\nu_{yy}}, \quad P_{xy}^*=\frac{\Omega_{xy}-a^* P_{yy}^*}{\nu_{yy}},
\eeq
where
\beq
\label{3.32}
\nu_{yy}=2\xi^*+\nu^{*\text{M}}\nu_\eta^*+2\mu \left(\frac{1+\theta}{\theta}\right)^{1/2} Y, 
\eeq
\beq
\label{3.32.1}
\Omega_{yy}=-\nu^{*\text{M}}\chi \left(\zeta^*-\nu_\eta^*\right)-2\mu \left(\frac{1+\theta}{\theta}\right)^{1/2} \left(X+Z P_{g,yy}^*\right),
\eeq
\beq
\label{3.32.2}
\Omega_{xy}=-2 \mu \left(\frac{1+\theta}{\theta}\right)^{1/2} Z P_{g,xy}^*.
\eeq
Note that the elements of the pressure tensor $P_{g,yy}^*$ and $P_{g,xy}^*$ of the molecular gas must be expressed in terms of the (reduced) shear rate $a^*$ and the (reduced) thermostat parameter $\xi^*$. For doing it, one has to take into account the relationships between $\widetilde{a}$ and $\widetilde{\xi}$ with $a^*$ and $\xi^*$, respectively. They are given by $\widetilde{a}=(\gamma/\widetilde{\nu}_g^\text{M})a^*$ and $\widetilde{\xi}=(\gamma/\widetilde{\nu}_g^\text{M})\xi^*$, where
\beq
\label{3.33.0}
 \frac{\gamma}{\widetilde{\nu}_g^\text{M}}= \frac{d+2}{\sqrt{2}d}\left(\frac{\overline{\sigma}}{\sigma_g}\right)^{d-1}\frac{m_g}{m}.
\eeq

The equation defining the temperature ratio $\chi$ can be easily derived from Eq.\ \eqref{3.26} as
\beq
\label{3.33}
\frac{2}{d}a^* P_{xy}^*+2\xi^* \chi=-\nu^{*\text{M}}\chi \zeta^*+2\mu^2 \left(\frac{1+\theta}{\theta}\right)^{1/2}\left(1-\chi\right).
\eeq
From Eqs.\ \eqref{3.31} and \eqref{3.33}, one gets finally $a^*$ in terms of the parameter space of the system:
\beq
\label{3.34}
a^*=\sqrt{\frac{d}{2}\frac{2\mu^2\left(1+\theta^{-1}\right)^{1/2}(1-\chi)-\left(\nu^{*\text{M}}\zeta^*+2\xi^*\right)\chi}{\frac{\Omega_{xy}/a^*}{\nu_{yy}}-\frac{\Omega_{yy}}{\nu_{yy}^2}}}.
\eeq
As happens in the case of IHS \cite{GChG23}, the temperature ratio $\chi$ cannot be expressed in Eq.\ \eqref{3.34} as an explicit function of the (reduced) shear rate and the remaining parameters of the system. On the other hand, for given values of the parameter space $\Xi\equiv \left(\chi, \al, \sigma/\sigma_g, m/m_g, \phi, T_g^*\right)$, the temperature ratio can be implicitly determined from the physical solution to Eq.\ \eqref{3.34}.

\subsection{Brownian limit}

Before illustrating the shear-rate dependence of the rheological properties of the molecular gas for arbitrary values of the mass ratio $m/m_g$, it is convenient to check the consistency of the present results with those derived in Ref.\ \cite{GGG19a} for IMM by using the Fokker--Planck operator \eqref{3.24}. This consistency is expected to apply in the Brownian limit $m/m_g\to \infty$. In this limiting case, at a given value of the (reduced) shear rate $a^*$, $\theta\to \infty$,  $\gamma/\nu_g^\text{M} \propto m_g/m\to 0$, $\widetilde{a}\propto m_g/m\to 0$, $\widetilde{\xi}\propto \widetilde{a}^2\propto (m_g/m)^2\to 0$, and $\xi^*=\widetilde{\xi}(\widetilde{\nu}_g^\text{M}/\gamma)\propto m_g/m\to 0$. Consequently, $P_{g,ij}^*=\delta_{ij}$ and the expressions of $P_{yy}^*$, $P_{xy}^*$, and $a^*$ are
\beq
\label{3.35}
P_{yy}^*=\frac{2-\nu^{*\text{M}} \chi \left(\zeta^*-\nu_\eta^*\right)}{2+\nu^{*\text{M}} \nu_\eta^*}, \quad P_{xy}^*=-\frac{2-\nu^{*\text{M}} \chi \left(\zeta^*-\nu_\eta^*\right)}{\left(2+\nu^{*\text{M}} \nu_\eta^*\right)^2}a^*,
\eeq
%=-\frac{a^*}{2+\nu^* \nu_\eta^*}P_{yy}^*
%\beq
%\label{41}
%P_{xy}^*=-\frac{a^*}{2+\nu^* \nu_\eta^*}P_{yy}^*=-\frac{2-\nu^* \chi \left(\zeta_0^*-\nu_\eta^*\right)}{\left(2+\nu^* \nu_\eta^*\right)^2}a^*,
%\eeq
\beq
\label{3.36}
a^*=\sqrt{\frac{d}{2}\frac{\nu^{*\text{M}} \zeta^*+2\left(1-\chi^{-1}\right)}{\nu^{*\text{M}} \left(\nu_\eta^*-\zeta^*\right)+2\chi^{-1}}}\left(2+\nu^{*\text{M}}\nu_\eta^*\right).
\eeq
Equations \eqref{3.35} and \eqref{3.36} are consistent with Eqs.\ (32), (33), and (35) of Ref.\ \cite{GGG19a}. It is important to note that, to assess consistency with the Fokker--Planck results, the size ratio has been kept constant or proportional to the mass ratio so that $\xi^*\to 0$.

\section{Rheology from a BGK-type kinetic model of the Boltzmann equation}
\label{sec4}

We consider in this Section the results derived for the USF from a BGK-type kinetic model of the Boltzmann equation \cite{BDS99,VGS07}. In the problem for the granular suspension considered here, one has to replace the true Boltzmann operators $J_g[f_g,f_g]$, $J[f,f]$, and $J_\text{BL}[f,f_g]$ by simpler relaxation terms that retain the relevant physical properties of those operators but are more tractable than the true kinetic equations. As in the case of IMM, let us determine separately the rheological properties of the molecular and granular gases by starting from these kinetic models.  

\subsection{Rheological properties of the molecular gas}

In the case of the molecular gas, the Boltzmann collision operator $J_g[f_g,f_g]$ is replaced by the conventional BGK kinetic model \cite{BGK54,C88}:   
\beq
\label{4.1}
J_g[f_g,f_g] \to -\nu_g\left(f_g-f_g^\text{M}\right),
\eeq
where $\nu_g$ is an effective velocity-independent collision frequency and $f_g^\text{M}$ is the Maxwellian distribution 
\beq
\label{4.2}
f_g^\text{M}(\mathbf{V})=n_g\left(\frac{m_g}{2\pi T_g}\right)^{d/2}\exp\left(-\frac{m_g V^2}{2T_g}\right).
\eeq
Thus, according to Eq.\ \eqref{2.8}, the velocity distribution function $f_g(\mathbf{V})$ obeys the BGK kinetic equation  
\beq
\label{4.3}
-a V_y \frac{\partial f_g}{\partial V_x}-\xi \frac{\partial}{\partial \mathbf{V}}\cdot \mathbf{V} f_g=-\nu_g\left(f_g-f_g^\text{M}\right).
\eeq
The nonzero elements of the pressure tensor $P_{g,ij}$ can be easily obtained from Eq.\ \eqref{4.3} by multiplying both sides of this equation by $m_g V_i V_j$ and integrating over $\mathbf{V}$. The BGK expressions of the (reduced) elements of the pressure tensor $P_{g,ij}^*$ are given by Eqs.\ \eqref{3.10} with the replacement $\widetilde{\nu}_g^\text{M}\to \nu_g$. As a consequence, the results derived from the BGK equation for the rheological properties agree with those obtained from the Boltzmann equation for IHS when 
\beq
\label{4.4}
\nu_g=\widetilde{\nu}_g^\text{M}=\frac{8\pi^{(d-1)/2}}{(d+2)\Gamma\left(\frac{d}{2}\right)}n_g \sigma_g^{d-1}\sqrt{\frac{T_g}{m_g}}.
\eeq

\subsection{Rheological properties of the granular gas}

In the case of the molecular gas, we consider the kinetic model proposed by Vega Reyes \emph{et al.} \cite{VGS07} for granular mixtures where the relaxation terms for the collision operators $J[f,f]$ and $J_\text{BL}[f,f_g]$ are defined, respectively, as  
\beq
\label{4.5}
J[f,f] \to -\nu'(\al)\left(f-f^\text{M}\right)+\frac{\epsilon}{2}\frac{\partial}{\partial \mathbf{V}}\cdot \mathbf{V}f,
\eeq
\beq
\label{4.6}
J_\text{BL}[f,f_g] \to -\nu\left(f-\widetilde{f}_g\right).
\eeq
In Eqs.\ \eqref{4.5} and \eqref{4.6},  
\beq
\label{4.6.1}
f^\text{M}(\mathbf{V})=n \left(\frac{m}{2\pi T}\right)^{d/2} \exp\left(-\frac{mV^2}{2T}\right)
\eeq
is the Maxwellian distribution of the granular gas,
\beq
\label{4.7}
\epsilon=\frac{2\pi^{(d-1)/2}}{d\Gamma\left(\frac{d}{2}\right)}n\sigma^{d-1}\sqrt{\frac{T}{m}}(1-\al^2),
\eeq
and the reference distribution function $\widetilde{f}_g(\mathbf{V})$ is taken to be the same as in the well-known Gross and Krook (GK) model for molecular (elastic) mixtures \cite{GK56}, i.e.,
\beq
\label{4.8}
\widetilde{f}_g(\mathbf{V})=n\left(\frac{m}{2\pi\tilde{T}}\right)^{d/2}\exp\left(-\frac{m V^2}{2\widetilde{T}}\right).
\eeq  
In Eqs.\ \eqref{4.5} and \eqref{4.6}, the quantities $\nu'$, $\nu$, and $\widetilde{T}$ are chosen to optimize the agreement with some properties of interest of the Boltzmann equation for IHS. The usual way of obtaining the above parameters is to ensure that the kinetic model reproduces the collisional transfer equations of momentum and energy for elastic collisions ($\alpha=1$). However, since $\mathbf{U}=\mathbf{U}_g$ in the USF, we only have one constraint (the one associated with the transfer of energy) instead of two, so that $\widetilde{T}$ and $\nu$ admit more than one form. Here, we propose a choice (see Appendix \ref{appA} for more  technical details) that leads to an excellent agreement with the results obtained for IHS from Grad's moment method \cite{GChG23}. More specifically, we take the following values of $\widetilde{T}$ and $\nu$:  
\beq
\label{4.9}
\widetilde{T}=T_g, \quad \nu=\frac{8\pi^{(d-1)/2}}{d\Gamma\left(\frac{d}{2}\right)}n_g\overline{\sigma}^{d-1}
 \frac{m m_g}{(m+m_g)^2}\left(\frac{2T_g}{m_g}+\frac{2T}{m}\right)^{1/2}.
\eeq
It still remains to completely define the model to chose the effective collision frequency $\nu'$. It is defined here to reproduce the collisional moment  
\beq
\label{4.10}
\int d\mathbf{v}\; m V_i V_j J[f,f]
\eeq
of the Boltzmann equation for IHS when one takes Grad's trial distribution for $f$ \cite{GChG23}. This leads to the expression 
\beq
\label{4.11}
\nu'=\frac{2\pi^{(d-1)/2}}{d(d+2)\Gamma\left(\frac{d}{2}\right)}n\sigma^{d-1}\sqrt{\frac{T}{m}}\left(1+\al\right)
\left[d+1+\left(d-1\right)\al\right].
\eeq

Therefore, the BGK kinetic equation for the sheared granular gas is given by 
\beq
\label{4.12}
-a V_y \frac{\partial f}{\partial V_x}-\xi \frac{\partial}{\partial \mathbf{V}}\cdot \mathbf{V} f=-\nu'\left(f-f^\text{M}\right)+\frac{\epsilon}{2}\frac{\partial}{\partial \mathbf{V}}\cdot \mathbf{V}f-\nu\left(f-\widetilde{f}_g\right),
\eeq
where $\nu$ and $\nu'$ are defined by Eqs.\ \eqref{4.9} and \eqref{4.11}, respectively, and the Maxwellian distribution $\widetilde{f}_g$ is given by Eq.\ \eqref{4.8} with $\widetilde{T}=T_g$.

The possibility of determining all the velocity moments of the distribution function is likely one of the main advantages of employing a kinetic model instead of the true Boltzmann equation. In the USF problem, it is convenient to define the general velocity moments
\beq
\label{4.13}
M_{k_1,k_2,k_3} =\int d\mathbf{V} V^{k_1}_xV_y^{k_2}V_z^{k_3}f(\mathbf{V}).
\eeq 
Although we are here essentially interested in the three-dimensional case, we will compute the velocity moments for $d=3$ and $d=2$. 
Note that for hard disks ($d=2$), $k_3=0$ in Eq.\ \eqref{4.13} since the $z$-axis is meaningless. The hierarchy of moment equations can be obtained by multiplying Eq.\ \eqref{4.12} by $V^{k_1}_xV_y^{k_2}V_z^{k_3}$ and integrating over $\mathbf{V}$. The result is
\beq
\label{4.14}
a k_1M_{k_1-1,k_2+1,k_3}+(\nu'+\nu+k\lambda)M_{k_1,k_2,k_3}=N_{k_1,k_2,k_3},
\eeq
where $\lambda=\xi+\epsilon/2$, $k=k_1+k_2+k_3$, and   
\beq
\label{4.15}
N_{k_1,k_2,k_3}=n \left(\frac{2T_g}{m}\right)^{k/2}\Big(\nu+\chi^{k/2} \nu' \Big)M_{k_1,k_2,k_3}^L.
\eeq
In Eq.\ \eqref{4.14}, for hard spheres ($d=3$)
\beq
\label{4.15.1}
M_{k_1,k_2,k_3}^L=\int d\mathbf{c}\; c_x^{k_1}c_y^{k_2}c_z^{k_3}e^{-c^2}=\pi^{-3/2}\Gamma\left(\frac{k_1+1}{2}\right)\Gamma\left(\frac{k_2+1}{2}\right)\Gamma\left(\frac{k_3+1}{2}\right),
\eeq
if $k_1$, $k_2$, and $k_3$ are even, being zero otherwise. For hard disks ($d=2$),  
\beq
\label{4.15.2}
M_{k_1,k_2,0}^L=
\pi^{-1}\Gamma\left(\frac{k_1+1}{2}\right)\Gamma\left(\frac{k_2+1}{2}\right)
\eeq
if $k_1$ and $k_2$ are even, being zero otherwise.

The solution to Eq.\ \eqref{4.14} can be written as (see Appendix B of Ref.\ \cite{GGG19} for some details)
\beq
\label{4.16}
M_{k_1,k_2,k_3}=\sum_{q=0}^{k_1}\frac{k_1!}{(k_1-q)!}\frac{(-a)^q}{(\nu'+\nu+k\lambda)^{1+q}}N_{k_1-q,k_2+q,k_3}.
\eeq
The nonzero elements of the pressure tensor $P_{ij}$ can be easily obtained from Eq.\ \eqref{4.16}. In dimensionless form, the BGK-expressions of the elements of $P_{ij}^*=P_{ij}/nT_g$ are
\beq
\label{4.17}
P_{yy}^*=P_{zz}^*=\frac{\nu+\nu' \chi}{\nu'+\nu+2\lambda}, \quad
P_{xy}^*=-\frac{\nu+\nu' \chi}{\left(\nu'+\nu+2\lambda\right)^2}a,
\eeq
\beq
\label{4.17.1}
P_{xx}^*=\frac{\nu+\nu' \chi}{\nu'+\nu+2\lambda}\Bigg[1+\frac{2a^2}{\left(\nu'+\nu+2\lambda\right)^2}\Bigg].
\eeq
The (steady) temperature ratio $\chi=T/T_g$ can be obtained from the constraint $P_{xx}^*+(d-1)P_{yy}^*=d\chi$. This yields the implicit equation
\beq
\label{4.18}
a^*=\left(\nu^{'*}+\nu^*+2\lambda^*\right)\sqrt{\frac{d}{2}\chi
\frac{\nu^{*}\left(1-\chi^{-1}\right)+2\lambda^*}{\nu^*+\nu^{'*}\chi}},
\eeq
where $a^*=a/\gamma$,
\beq
\label{4.19}
\nu^{'*}=\frac{\nu'}{\gamma}=\frac{2^d}{d+2}\phi \sqrt{\frac{\chi T_g^*}{\pi}}(1+\al)\left[d+1+\left(d-1\right)\al\right],
\eeq
\beq
\label{4.20}
\nu^*=\frac{\nu}{\gamma}=2\mu^2 \left(\frac{1+\theta}{\theta}\right)^{1/2}, \quad
\lambda^*=\frac{\lambda}{\gamma}=\xi^*+\frac{\epsilon^*}{2},
\eeq
\beq
\label{4.21}
\xi^*=\frac{\sqrt{2}d}{d+2}\left(\frac{\sigma_g}{\overline{\sigma}}\right)^{d-1}\frac{m}{m_g}\widetilde{\xi}, \quad \epsilon^*=2^d\phi \sqrt{\frac{\chi T_g^*}{\pi}}(1-\al^2).
\eeq
Here, $\widetilde{\xi}$ is given by Eq.\ \eqref{3.13} and we recall that $T_g^*=T_g/m\sigma^2\gamma^2$.

\subsection{Brownian limit}

As in the case of IMM, it is quite interesting to consider the Brownian limiting case $m/m_g\to \infty$. In this case, $P_{g,ij}^*=\delta_{ij}$, $\mu\to 1$, $\theta\to \infty$, and $\xi^*\to 0$. Thus, following similar steps as those made for IMM, one gets the expressions   
\beq
\label{4.22}
P_{yy}^*=\frac{2+\nu^{'*}\chi}{2+\nu^{'*}+\epsilon^*}, \quad P_{xy}^*=-\frac{2+\nu^{'*}\chi}{(2+\nu^{'*}+\epsilon^*)^2}a^*, 
\eeq
\beq
\label{4.23}
a^*=\left(2+\nu^{'*}+\epsilon^*\right)\sqrt{\frac{d}{2}\chi
\frac{2\left(1-\chi^{-1}\right)+\epsilon^*}{2+\nu^{'*}\chi}}.
\eeq

\subsection{Velocity distribution of the granular gas}

Apart from obtaining all the velocity moments, the use of kinetic models allow us in some cases to explicitly determine the velocity distribution function $f(\mathbf{V})$. The BGK-type equation \eqref{4.12} reads
\beq
\label{4.24}
\left(1-d\widehat{\lambda}-\widehat{a}V_y \frac{\partial}{\partial V_x}-
\widehat{\lambda}\mathbf{V}\cdot \frac{\partial}{\partial \mathbf{V}}\right)f(\mathbf{V})=f_\text{R}(\mathbf{V}),
\eeq
where $\widehat{a}=a/(\nu'+\nu)$, $\widehat{\lambda}=\lambda/(\nu'+\nu)$, and 
\beq
\label{4.25}
f_\text{R}(\mathbf{V})=\frac{\nu'}{\nu'+\nu}f^\text{M}(\mathbf{V})+\frac{\nu}{\nu'+\nu}\widetilde{f}_g(\mathbf{V}).
\eeq
The hydrodynamic solution to Eq.\ \eqref{4.24} can be formally written as
\beqa
\label{4.26}
f(\mathbf{V})&=&\Big(1-d\widehat{\lambda}-\widehat{a}V_y \frac{\partial}{\partial V_x}-
\widehat{\lambda}\mathbf{V}\cdot \frac{\partial}{\partial \mathbf{V}}\Big)^{-1}f_\text{R}(\mathbf{V}) \nonumber\\
&=&\int_0^\infty ds\; e^{-(1-d\widehat{\lambda})s}e^{\widehat{a}sV_y \frac{\partial}{\partial V_x}}e^{\widehat{\lambda}s\mathbf{V}\cdot \frac{\partial}{\partial \mathbf{V}}}f_\text{R}(\mathbf{V}).
\eeqa
The action of the velocity operators in Eq.\ \eqref{4.26} on an arbitrary function $g(\mathbf{V})$ is \cite{GGG19}
\beq
\label{4.27}
e^{\widehat{a}sV_y \frac{\partial}{\partial V_x}}g(V_x,V_y,V_z)=g(V_x+\widehat{a}sV_y,V_y,V_z), 
\eeq
\beq
\label{4.28}
e^{\widehat{\lambda}s\mathbf{V}\cdot \frac{\partial}{\partial \mathbf{V}}}g(V_x,V_y,V_z)=g\left(e^{\widehat{\lambda}s}V_x, e^{\widehat{\lambda}s}V_y, e^{\widehat{\lambda}s}V_z\right).
\eeq
The explicit form of the one-particle velocity distribution function can be explicitly obtained when one takes into account in Eq.\ \eqref{4.26} the action of the velocity operators given by Eqs.\ \eqref{4.27} and \eqref{4.28}. The result can be written as 
\beq
\label{4.29}
f(\mathbf{V})=n\left(\frac{m}{2T}\right)^{d/2}\varphi(\mathbf{c}), \quad \mathbf{c}=\left(\frac{m}{2T}\right)^{1/2}\mathbf{V},
\eeq
where 
\beqa
\label{4.30}
\varphi(\mathbf{c})&=&\pi^{-d/2}\int_0^\infty ds\;e^{-(1-d\widehat{\lambda})s}\Bigg\{ \frac{\nu'}{\nu'+\nu}\exp\Big[-e^{2\widehat{\lambda}s}\Big(\mathbf{c}+
s\widehat{\boldsymbol{a}}
\cdot \mathbf{c}\Big)^2\Big]\nonumber\\
& & +\frac{\nu}{\nu'+\nu}\chi^{d/2}\exp\Big[-\chi e^{2\widehat{\lambda}s}\Big(\mathbf{c}+s\widehat{\boldsymbol{a}}\cdot \mathbf{c}\Big)^2\Big]\Bigg\}.
\eeqa
Here, we have introduced the tensor $\widehat{a}_{ij}=\widehat{a}\delta_{ix}\delta_{jy}$.

In order to illustrate the shear-rate dependence of the distribution function, it is convenient to consider the marginal distribution for $d=3$: 
\beqa
\label{4.31}
\varphi_x(c_x)&=&\int_{-\infty}^{\infty}\; dc_y \int_{-\infty}^{\infty}\; dc_z\; \varphi(\mathbf{c}) \nonumber\\
&=&\frac{1}{\sqrt{\pi}}\int_0^\infty\; ds\; \frac{e^{-(1-\widehat{\lambda})s}}{\sqrt{1+\widehat{a}^2s^2}}\Bigg\{
\frac{\nu'}{\nu'+\nu}\exp\Big(-e^{2\widehat{\lambda}s}\frac{c_x^2}
{1+\widehat{a}^2s^2}\Big)\nonumber\\
& & +\frac{\nu}{\nu'+\nu}\chi^{1/2}\exp\Big(-\chi e^{2 \widehat{\lambda}s}\frac{c_x^2}
{1+\widehat{a}^2s^2}\Big)\Bigg\}.
\eeqa
In the Brownian limit, $\xi^*\to 0$, $\mu\to 1$, $\theta\to \infty$, and so $\nu^*\to 2$, $\lambda^*\to \epsilon^*/2$, and 
\beq
\label{4.32}
\widehat{\lambda}\to \frac{\epsilon^*/2}{2+\nu^{'*}}, \quad \widehat{a}\to \frac{a^*}{2+\nu^{'*}}.
\eeq
Thus, when $m/m_g\to \infty$, Eq.\ \eqref{4.31} becomes
\beqa
\label{4.33}
\varphi_x(c_x)
&=&\frac{1}{\sqrt{\pi}}\int_0^\infty\; ds\; \frac{e^{-(1-\widehat{\lambda})s}}{\sqrt{1+\widehat{a}^2s^2}}\Bigg\{
\frac{\nu^{'*}}{2+\nu^{'*}}\exp\Big(-e^{2\widehat{\lambda}s}\frac{c_x^2}
{1+\widehat{a}^2s^2}\Big)\nonumber\\
& & +\frac{2}{2+\nu^{'*}}\chi^{1/2}\exp\Big(-\chi e^{2\widehat{\lambda}s}\frac{c_x^2}
{1+\widehat{a}^2s^2}\Big)\Bigg\},
\eeqa
where $\widehat{\lambda}$ and $\widehat{a}$ are given by Eqs.\ \eqref{4.32}.

\section{Comparison between IMM and BGK results}
\label{sec5}

In Sections \ref{sec3} and \ref{sec4}, we made use of the Boltzmann equation for IMM and the BGK-type kinetic model to investigate the shear-rate dependence of rheological properties in a sheared granular suspension. These properties are expressed in terms of the coefficient of restitution $\alpha$, the reduced background temperature $T_g^*$, and the diameter $\sigma/\sigma_g$ and mass $m/m_g$ ratios. Additionally, there exists a residual dependence on density through the volume fraction $\phi$. To avoid that, one could for instance reduce the shear rate using the effective collision frequencies $\nu^\text{M}(T)$ or $\nu(T)$. However, for consistency with simulations and considering the background temperature $T_g$ as a known quantity, we opted to employ $\gamma(T_g)$ as the reference frequency.

Given that in this Section the second-degree moments of the distribution function are compared with molecular dynamics (MD) simulations for IHS in the Brownian limiting case \cite{HTG17}, we set fixed values of $T_g^*=1$ and $\phi=0.0052$ for subsequent analysis. The selection of $T_g^*$ as a free parameter imposes a constraint between the diameter $\sigma/\sigma_g$ and mass $m/m_g$ ratios \cite{GChG23}:
\begin{equation}
\label{5.1}
\frac{\sigma}{\sigma_g}=\left[\left(\frac{\sqrt{\pi}}{4\sqrt{2}}\frac{n}{n_g}\sqrt{\frac{m}{m_g}}\frac{1}{\phi\sqrt{T_g^*}}\right)^{1/(d-1)}-1\right]^{-1}.
\end{equation}
This relation ensures convergence of results to those obtained via the Fokker--Planck equation as $m/m_g\to \infty$ since $\xi^*\to 0$. Furthermore, since we want to recover the results obtained in Ref.\ \cite{GChG23} derived from Grad's method, we take $n/n_g=10^{-3}$ (rarefied granular gas).

\begin{figure}[h!]
\begin{center}
\includegraphics[width=1\columnwidth]{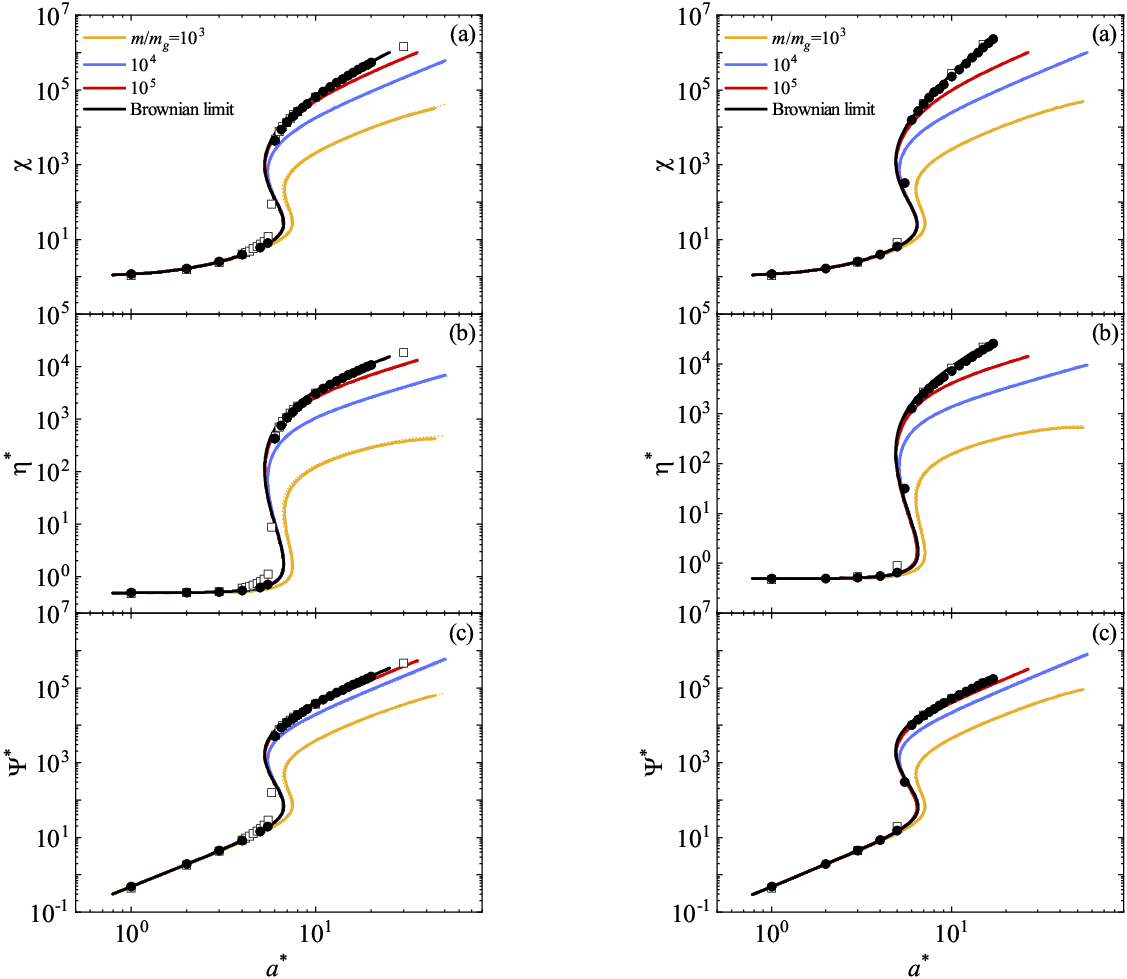}
\end{center}
\caption{Plots of the (steady) granular temperature $\chi$ (a), the non-Newtonian shear viscosity $\eta^*$ (b), and the normal stress difference $\Psi^*$ (c) as functions of the (reduced) shear rate $a^*$  for two different values of the coefficient of restitution $\alpha$: 0.9 (left panel) and 1 (right panel). The graphs represent four distinct mass ratio values $m/m_g$: $10^3$ (yellow lines), $10^4$ (blue lines), $10^6$ (red lines), and the Brownian limit (black lines). Here, $T_g^*=1$, $d=3$, and $\phi=0.0052$. The solid lines correspond to the IMM results, the dashed lines are the BGK-like results, and the 
dotted lines refer to Grad's solution for IHS. Symbols denote computer simulation results performed in the Brownian limit: circles refer to the DSMC data obtained in this paper for IHS while squares are MD results obtained in Ref.\ \cite{HTG17} for IHS. 
\label{fig1}}
\end{figure}

The second-degree moments expressed through the reduced temperature $\chi$, non-Newtonian shear viscosity $\eta^*$, and the normal stress difference $\Psi^*$ are plotted in Fig.\ \ref{fig1} for $\alpha=0.9$ and $1$. Here, $\Psi^*=P_{xx}^*-P^*_{yy}=d\chi-dP^*_{yy}$. Equations \eqref{3.31} and \eqref{4.17} provide analytical expressions for IMM and BGK-type kinetic model, respectively, from which rheological properties are illustrated. Notably, there is nearly perfect agreement between Grad's solution for IHS as obtained in Ref.\ \cite{GChG23} and both IMM and BGK-type results for any mass ratio, highlighting the ability of relatively simple models to capture essential properties of granular suspensions.

In particular, a DST transition characterized by $S$-shaped curves becomes more pronounced as the mass ratio $m/m_g$ increases. Specifically, the non-Newtonian shear viscosity $\eta^*$ exhibits a discontinuous transition (at a certain value of $a^*$) which intensifies as the particles of the granular gas become heavier. The theoretical results are validated with MD simulations \cite{HTG17}  in the Brownian limiting case ($m/m_g\to 0$), showing generally good agreement despite slight discrepancies in the transition zone. Simulations suggest a sharper transition, likely due to the absence of molecular chaos in highly non-equilibrium situations. To address this, DSMC simulations for IHS are performed in the same limit, showing good agreement with theoretical results and further emphasizing a more pronounced transition. This phenomenon is likely attributable to a sudden growth of higher-order moments causing the distribution function to strongly deviate from the reference Maxwellian approximation \eqref{4.6.1}. Some technical details of the application of the DSMC method are available in the supplementary material of Ref.\ \cite{GG22a} [\url{https://doi.org/10.1017/jfm.
2022.410}]. 

The simplicity of the BGK and IMM models enables exploration beyond second-degree moments. Accordingly, we utilize the BGK-type kinetic equation to compute the fourth-degree moments. Although similar calculations could be performed in the case of IMM, we opt to omit them due to their extensive analytical effort. Additionally, drawing insights from the Fokker--Planck model \cite{GGG19a} and dry granular gases \cite{SG07}, we anticipate potential divergences of the moments derived from IMM under certain shear rate conditions. We focus our efforts on calculating the following fourth-degree moments
\beq
\label{5.2}
M^*_{4|0}=\frac{m^2}{n T_g^2}\int d{\bf V} V^4 f(\mathbf{V}),
\eeq
\beq
\label{5.2.1}
M_{2|xy}^*=\frac{m^2}{n T_g^2}\int d{\bf V} V^2 V_xV_y f(\mathbf{V}).
\eeq
Thus, in terms of the generic moments $M_{k_1,k_2, k_3}$ and 
according to the expression \eqref{4.16}, the moments $M^*_{4|0}$ and $M_{2|xy}^*$ are given by  
\beq
\label{5.3}
M^*_{4|0}=\frac{m^2}{n T_g^2}\left(M_{4,0,0}+M_{0,4,0}+M_{0,0,4}+2M_{2,2,0}+2M_{2,0,2}+2M_{0,2,2}\right),
\eeq
\beq
\label{5.4}
M^*_{2|xy}=\frac{m^2}{n T_g^2}\left(M_{3,1,0}+M_{1,3,0}+M_{1,1,2}\right).
\eeq

\begin{figure}[h!]
\begin{center}
\includegraphics[width=1\columnwidth]{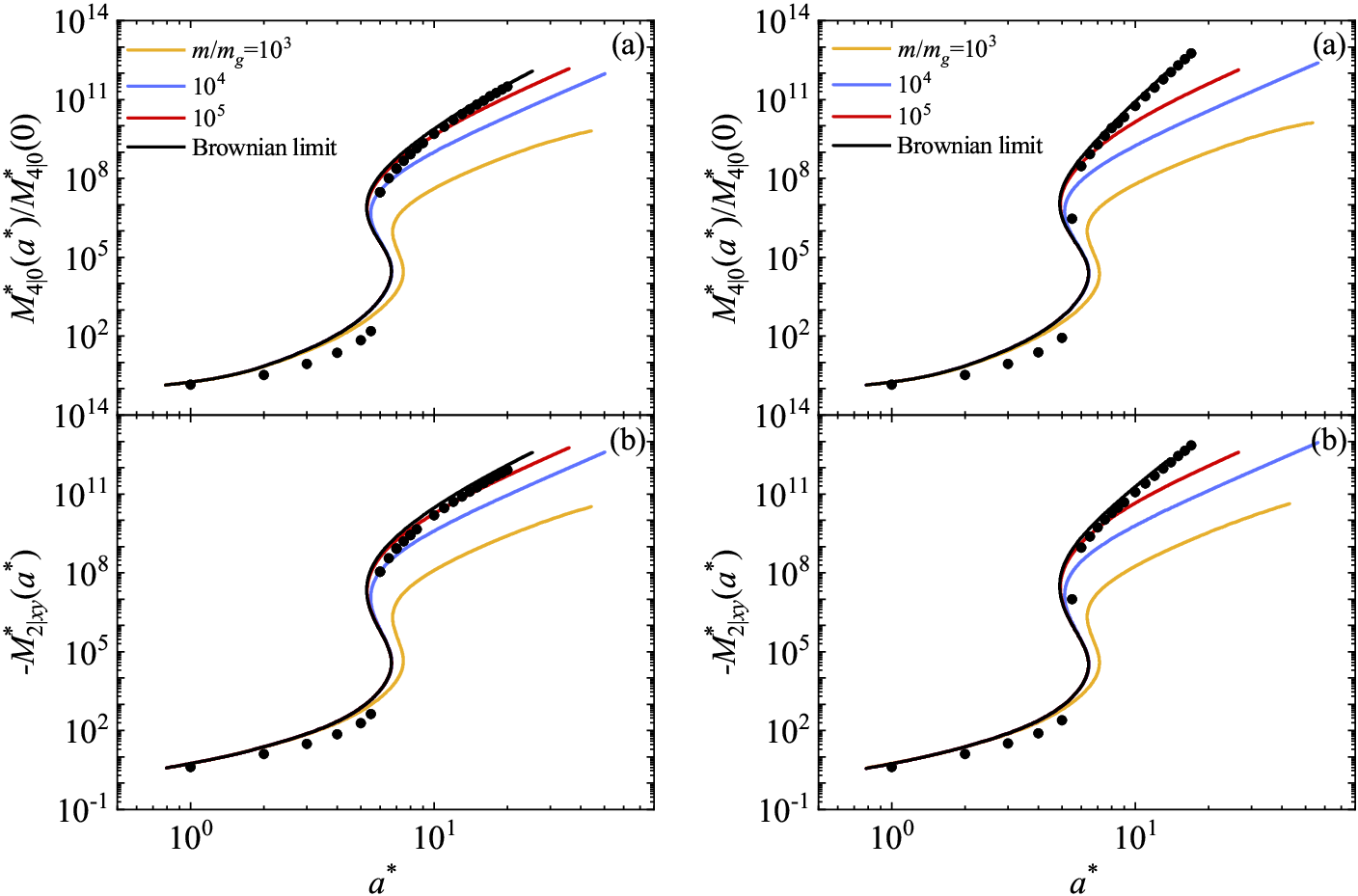}
\end{center}
\caption{Plots of the (reduced) fourth-degree moments $M_{4|0}^*(a^*)/M_{4|0}^*(0)$ (a) and $-M_{2|xy}^*(a^*)$ (b) as functions of the (reduced) shear rate $a^*$ obtained from the BGK-type equation for two different values of the coefficient of restitution $\alpha$: 0.9 (left panel) and 1 (right panel). The graphs represent four distinct mass ratio values $m/m_g$: $10^3$ (yellow lines), $10^4$ (blue lines), $10^6$ (red lines), and the Brownian limit (black lines). Here, $T_g^*=1$, $d=3$, and $\phi=0.0052$. Symbols refer to the DSMC results obtained for IHS in this paper in the Brownian limit. 
\label{fig2}}
\end{figure}

Figure \ref{fig2}(a) illustrates the ratio $M^*_{4|0}(a^*)/M^*_{4|0}(0)$ as a function of $a^*$ for $\alpha=0.9$ and $1$ and three values of the mass ratio. We observe that variations in the mass ratio do not significantly alter the trends observed in the Brownian limiting case \cite{GGG19a}. An abrupt transition in the higher-order moments is evident within a small region of $a^*$. Specifically, the kurtosis $M^*_{4|0}$ increases with the mass ratio $m/m_g$ until it converges to the value obtained in the Brownian limit. Consistent with the conclusions drawn in Ref. \cite{GChG23}, an increase in the mass of the particles of the granular gas results in an elevation of the granular temperature. Consequently, non-equipartition accentuates and moves the suspension away from equilibrium, leading to an increase in kurtosis as the distribution function deviates from its Maxwellian form. Regarding the influence of collisional dissipation, we observe that the effect of $\alpha$ on $M^*_{4|0}$ remains relatively discrete. Figure \ref{fig2}(b) illustrates the shear-rate dependence of the (reduced) moment $M_{2|xy}^*$. This moment vanishes in the absence of shear rate ($a^*=0$). Similar conclusions to those made for the moment $M^*_{4|0}$ can be drawn.

\begin{figure}[h!]
\begin{center}
\includegraphics[width=0.5\columnwidth]{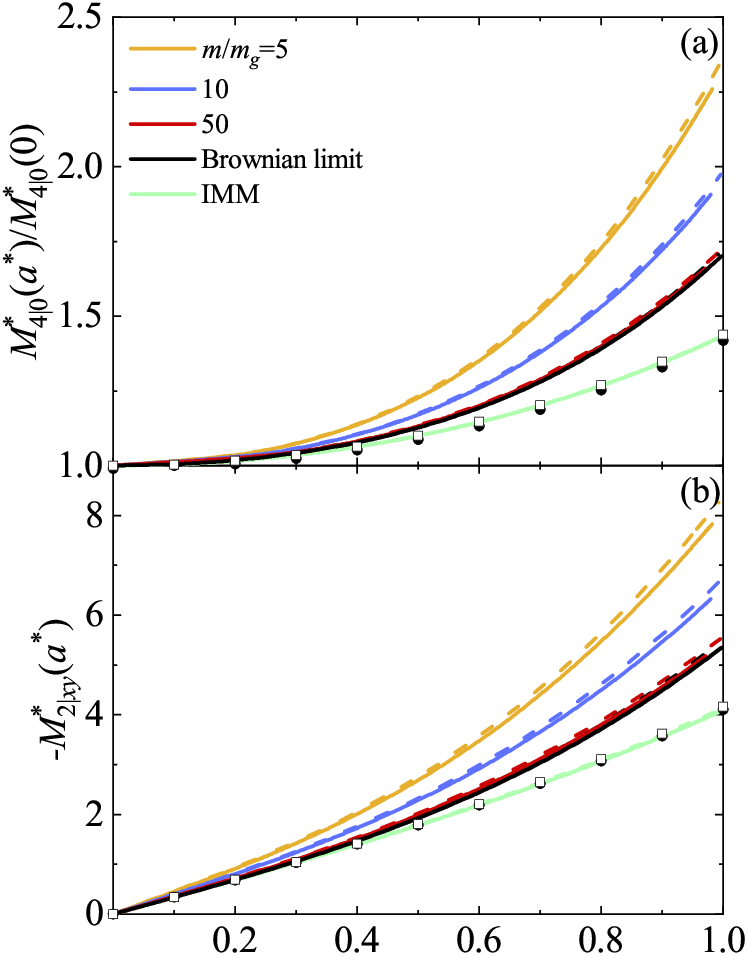}
\end{center}
\caption{Plots of the (reduced) fourth-degree moments $M_{4|0}^*(a^*)/M_{4|0}^*(0)$ (a) and $-M_{2|xy}^*(a^*)$  (b) as functions of the (reduced) shear rate $a^*$ obtained from the BGK-type equation for two different values of the coefficient of restitution $\alpha$: 0.7 (solid lines) and 1 (dashed lines). The graphs represent four distinct mass ratio values $m/m_g$: $5$ (yellow lines), $10$ (blue lines), $50$ (red lines), and the Brownian limit (black lines). Here, $T_g^*=1$, $d=3$, and $\phi=0.0052$. Symbols refer to the DSMC results for IHS in the Brownian limit (squares for $\alpha=1$ and circles for $\alpha=0.7$). The green lines are the IMM results as obtained in Ref.\ \cite{GGG19} in the Brownian limit.
\label{fig3}}
\end{figure}

Theoretical predictions for the fourth-degree moments are compared against DSMC simulations for IHS conducted in this paper in the Brownian limiting case. A qualitative agreement is observed, although simulations suggest a sharper transition. Some quantitative discrepancies are noticeable, which are mainly disguised by the scale. To assess the reliability of the BGK-type results, we focus on the region $0 < a^* < 1$, where all the fourth-degree velocity moments of IMM are well-defined functions of the shear rate. In addition, non-Newtonian effects are still significant in the range of values of the (reduced) shear rate $a^* \leq 1$. To this purpose, Fig.\ \ref{fig3} shows the (reduced) fourth-degree moments $M_{4|0}^*(a^*)/M_{4|0}^*(0)$ and $M_{2|xy}^*(a^*)$ for $\alpha=0.7$ and $1$. These moments are also illustrated as obtained for IMM in the Brownian limit \cite{GGG19a}. It is worth noting that the results derived in Ref.\ \cite{GGG19a} stem from considering an effective force modeling the interstitial gas, diverging from the limit of a Boltzmann--Lorentz operator modeled by a BGK-type equation as $m/m_g\to \infty$. Consequently, since DSMC simulations employ the exact Fokker--Planck operator, they perfectly align with the IMM results, while discrepancies emerge when compared with the BGK-type results. It is noteworthy that the BGK-type model slightly overestimates the deviation from the Newtonian situation ($a^*=0$), a phenomenon also observed for molecular gases \cite{GS05}. Moreover, non-Newtonian effects become apparent even at low values of $a^*$.

\begin{figure}[h!]
\begin{center}
\includegraphics[width=0.5\columnwidth]{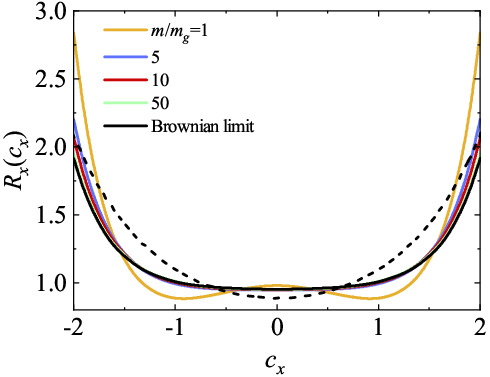}
\end{center}
\caption{Plot of the ratio $R_x(c_x) = \varphi_x(c_x)/(\pi^{-1/2}e^{-c_x^2})$ for $a^*=1$ as a function of the (reduced) velocity
$c_x$ for $\alpha=0.9$ and five different values of the mass ratio $m/m_g$: $1$ (yellow line), $5$ (blue line), $10$ (red line), $50$ (green line), and the Brownian limit (black lines). Here, $T_g^*=1$, $d=3$, and $\phi=0.0052$. The dashed line refer to the DSMC results for IHS in the Brownian limit.
\label{fig4}}
\end{figure}

Finally, in Fig.\ \ref{fig4}, the ratio $R_x(c_x) = \varphi_x(c_x)/(\pi^{-1/2}e^{-c_x^2})$ is plotted for $a^*=1$ and four different values of the mass ratio. Here, the marginal distribution $\varphi_x(c_x)$ is given by Eq.\ \eqref{4.31}. It is evident that the deviation from equilibrium ($R_x \neq 1$) becomes more significant as the mass ratio $m/m_g$ increases. Moreover, a comparison between theory and DSMC simulations reveals some disagreement in the BGK-type solution. %particularly in the region of thermal velocities ($|c_x| \sim 1$). 
Although the relative difference of these discrepancies is relatively small (it is about 8\%), this contradicts what has been observed in Ref. \cite{ChVG15}, where good agreement between the BGK solution and DSMC data is shown in the region of thermal velocities.

\section{Concluding remarks}
\label{sec6}

 In our study, we have explored the non-Newtonian transport properties of a dilute granular suspension subjected to USF using the Boltzmann kinetic equation. The particles are represented as $d$-dimensional hard spheres with mass $m$ and diameter $\sigma$, immersed in an interstitial gas acting as a thermostat at temperature $T_g$. Various models for granular suspensions incorporate a gas-solid force to represent the influence of the external fluid. While some models consider only isolated body resistance via a linear drag law \cite{TK95,SMTK96,PS12,H13,WGZS14,SA17,ASG19,SA20}, others \cite{GTSH12,PLMPV98} include an additional Langevin-type stochastic term. 
In this paper, we consider a suspension model where the collisions between grains and particles of the interstitial (molecular) gas are taken into account. Thus,
based on previous studies \cite{BMP02a,S03a}, we discretize the surrounding molecular gas, assigning individual particles with mass $m_g$ and diameter $\sigma_g$, thereby accounting for \emph{elastic} collisions between grains and background gas particles in the starting kinetic equation.

Under USF conditions, the system is characterized by constant density profiles $n$ and $n_g$, uniform temperatures $T$ and $T_g$, and a (common) flow velocity $U_x = U_{g,x}=ay$, where $a$ denotes the shear rate. In agreement with previous investigations on uniform sheared suspensions, the mean flow velocity $\mathbf{U}$ is coupled to that of the gas phase $\mathbf{U}_g$. Consequently, the viscous heating term due to shear and the energy transferred by the grains from collisions with the molecular gas are compensated by the cooling terms derived from collisional dissipation, allowing the achievement of a steady state. A distinctive feature of the USF is that the one-particle velocity distribution function $f(\mathbf{r},\mathbf{v})$ depends on space only through its dependence on the peculiar velocity $\mathbf{V}=\mathbf{v}-\mathbf{U}$. Consequently, the velocity distribution function becomes uniform in the Lagrangian reference frame moving with $\mathbf{V}$. This means that  $f(\mathbf{r},\mathbf{v})\equiv f(\mathbf{V})$. Based on symmetry considerations, the heat flux $\mathbf{q}$ vanishes, making the pressure tensor $\mathrm{P}$ the relevant flux. Therefore, to understand the intricate dynamics of granular suspensions under shear flow, it is imperative to focus on their non-Newtonian properties ---derived from the pressure tensor. These include the (reduced) temperature $\chi=T/T_g$, the (reduced) nonlinear shear viscosity $\eta^*$, and the (reduced) normal stress difference $\Psi^*$. 

Given that the most challenging aspect of dealing with the Boltzmann equation lies in the collision operator, it is reasonable to explore alternatives that render this operator more analytically tractable than in the case of IHS. Among the most sophisticated techniques in this regard is to consider the Boltzmann equation for IMM. As in the case of elastic collisions \cite{IT56,GS03}, the collision rate for IMM is independent of the relative velocity of the colliding particles. As a consequence, the collisional moments of degree $k$ of the Boltzmann collisional operator can be expressed as a linear combination of velocity moments of degree less than or equal to $k$. To complement the results derived for IMM, we have also considered in this paper the use of a BGK-type kinetic model where the true Boltzmann operator is replaced by a simple relaxation term. Here, we employed both approaches to compute the rheological properties of the sheared granular suspension. Thus, our objective was twofold. Firstly, we aimed to assess the reliability and compatibility of the proposed models with previous results \cite{GChG23} obtained for IHS using Grad's moment method. Additionally, DSMC simulations for IHS were performed as an alternative method to validate any potential discrepancies identified. Secondly, taking advantage of the capabilities provided by the BGK model, we endeavored to calculate the velocity distribution function and the higher-order moments that offer insights into its characteristics.

Before proceeding with the computation of rheological properties, it is necessary to understand the response of the molecular fluid to shear stress. This assessment was also conducted using both IMM and BGK-type kinetic model that were later used to model the granular gas. A novelty here is the application of a force (Gaussian thermostat) of the form $\mathbf{F}=-m\xi\mathbf{V}$ to compensate for the energy gained through viscous shear stresses. This force, by consistency, also applies to the granular gas, maintaining convergence to a steady state. As anticipated, the results agree well with those obtained through Grad's moment method \cite{GChG23} [see Eqs.\ \eqref{3.15} and \eqref{4.4}]. Consequently, once the problem conditions (including the shear rate $a$) are defined, the molecular temperature $T_g$ is determined, effectively serving as a thermostat for the granular gas.

After accurately describing the rheology of the molecular gas, we focused on modeling the granular gas. Using both IMM and BGK-type model separately, we have calculated the non-zero elements of the the pressure tensor. The knowledge of these elements allows us to identify the relevant rheological properties of the granular suspension. As shown in Fig.\ \ref{fig1}, these quantities are represented as functions of the coefficient of restitution $\alpha$ and the mass ratio $m/m_g$. In particular, we find that the theoretical results obtained from the Grad's method for IHS, IMM and BGK-type model show remarkable agreement, with almost indistinguishable curves. This underlines the effectiveness of \emph{structurally simple} models in capturing the complexities of sheared granular suspensions. We observe a DST-type transition starting at a certain value of $a^*$, which increases with the mass ratio $m/m_g$. 
%In addition, DSMC simulations for IHS were performed to validate the accuracy of the different approximations used in solving the Boltzmann equation in the Brownian limit. 
Interestingly, similar to the MD simulations performed for IHS \cite{HTG17}, the DSMC data suggest a more abrupt transition than predicted by theory. Given that the main divergences between Grad's (for IHS) and DSMC results arise from the form of the distribution function, the significance of investigating higher-order moments to assess the deviation of the distribution function from its Maxwellian reference is then justified.

Based on previous literature where discrepancies in fourth-degree moments have been observed \cite{GGG19,SG07}, and acknowledging the potential lengthiness of calculations, for the sake of simplicity, we decided to employ only the BGK-type model to compute higher-order moments. Specifically, we concentrated on the (symmetric) fourth-degree moments $M_{4|0}$ and $M_{2|xy}$. The shear-rate dependence of these moments is illustrated in Fig.\ \ref{fig2} for the same parameter values of $\alpha$ and $m/m_g$ as those employed for the rheological quantities. Initially, we note that the fourth-degree moments also exhibit an abrupt transition at a value of $a^*$ that increases with $m/m_g$ until reaching the Brownian limit. Furthermore, DSMC simulations in the Brownian limit qualitatively capture the profile of these moments, although some quantitative disparities are apparent. To ascertain the extent of these discrepancies, we narrowed our focus to the interval $0<a^*<1$ where non-Newtonian effects are apparent. Additionally, we included IMM results directly as obtained from an effective Fokker-Planck-type model \cite{GGG19a}. Figure \ref{fig3} illustrates that BGK results overestimate the deviation from the moments computed when no shear stress is applied compared to DSMC simulations and the results obtained using an effective force to model the interstitial gas. These disparities are also observed in the marginal distribution function $\phi_x$.

The theoretical findings presented here motivate the comparison with computer simulations. Although the observed agreement in the Brownian limit is encouraging, there is scope to extend this agreement to scenarios with finite mass ratios. Our plan is to carry out simulations of this type in the near future, which we expect will further validate and improve our theoretical framework. In addition, we plan to extend our current findings to finite densities by exploring the Enskog kinetic equation, which will allow us to evaluate the involvement of density in the occurrence of these phenomena. This line of research will be one of the main objectives of our upcoming research.

\authorcontributions{ ``Conceptualization, R. G. G. and V. G.; software, validation, R. G. G. and V. G.; formal analysis,  R. G. G. and V. G.;  writing---original draft preparation, V. G.; writing---review and editing,  R. G. G. and V. G. All authors have read and agreed to the published version of the manuscript.''}

\funding{The authors acknowledge financial support from Grant PID2020-112936GB-I00 funded by MCIN/AEI/ 10.13039/501100011033, and from Grant IB20079 funded by Junta de Extremadura (Spain) and by ERDF ``A way of making Europe.'' 
}

%\institutionalreview{Not applicable}

%\informedconsent{Not applicable}

%\dataavailability{The data that support the findings of this study are available from the corresponding author upon reasonable request.} 

%\acknowledgments{In this section you can acknowledge any support given which is not covered by the author contribution or funding sections. This may include administrative and technical support, or donations in kind (e.g., materials used for experiments).}

\conflictsofinterest{The authors report no conflict of interest.}

%\section*{Acknowledgments}

%The authors acknowledge financial support from Grant PID2020-112936GB-I00 funded by MCIN/AEI/ 10.13039/501100011033, and from Grant IB20079 funded by Junta de Extremadura (Spain) and by ERDF ``A way of making Europe.'' 

\appendixtitles{no} % Leave argument "no" if all appendix headings stay EMPTY (then no dot is printed after "Appendix A"). If the appendix sections contain a heading then change the argument to "yes".
%\appendixstart
\appendix
\section[\appendixname~\thesection]{Some technical details in the BGK-type kinetic model}
\label{appA}

In this Appendix we give some details on the determination of the parameters $\widetilde{T}$ and $\nu$ appearing in the relaxation term \eqref{4.6}. To get them, we require the collisional transfer of energy of grains due to their elastic collisions with particles of the molecular gas to be the same as the one obtained from the true Boltzmann--Lorentz collision operator. This implies that 
\beq
\label{a1}
\int d\mathbf{v} V^2 J_\text{BL}[\mathbf{V}|f,f_g]=-\nu\int d\mathbf{v} V^2 \left[f(\mathbf{V})-\widetilde{f}_g(\mathbf{V})\right].
\eeq
Given that the collisional moment  involving the operator $J_\text{BL}[f,f_g]$ cannot be exactly computer for IHS, one estimate this moment by replacing $f$ and $f_g$ by their Grad's solutions \cite{G49}. In this approximation, one achieves the results \cite{GChG23}
\beq
\label{a2}
\int d\mathbf{v} V^2 J_\text{BL}[\mathbf{V}|f,f_g]=-\frac{8\pi^{(d-1)/2}}{\Gamma\left(\frac{d}{2}\right)}n n_g\overline{\sigma}^{d-1}
 \frac{m_g}{(m+m_g)^2}\left(\frac{2T_g}{m_g}+\frac{2T}{m}\right)^{1/2}\left(T-{T}_g\right).
\eeq
Moreover, 
\beq
\label{a3}
\int d\mathbf{v} V^2 \left[f(\mathbf{V})-\widetilde{f}_g(\mathbf{V})\right]=-d \frac{n \nu}{m}\left(T-\widetilde{T}\right).
\eeq
From Eqs.\ \eqref{a2} and \eqref{a3} one gets the identity
\beq
\label{a4}
\frac{8\pi^{(d-1)/2}}{d\Gamma\left(\frac{d}{2}\right)} n_g\overline{\sigma}^{d-1}
 \frac{m m_g}{(m+m_g)^2}\left(\frac{2T_g}{m_g}+\frac{2T}{m}\right)^{1/2}\left(T-{T}_g\right)=\nu \left(T-\widetilde{T}\right).
\eeq
Equation \eqref{a4} allows to make the identifications \eqref{4.9}. 

Finally, from Grad's moment method \cite{G49}, the collisional moment \eqref{4.10} can be written as \cite{GChG23} 
\beq
\label{a5}
\int d\mathbf{v}\; m V_i V_j J[f,f]=-\nu^\text{M}  n T_g \left[\nu_\eta^{*\text{IHS}} P_{ij}^*-\chi\left(\nu_\eta^{*\text{IHS}}-\zeta^{*\text{IHS}}\right) \delta_{ij}\right],
\eeq
where $\nu^\text{M}$ is defined in Eq.\ \eqref{3.22} and 
\begin{equation}
\label{a6}
%\nu^{\text{IHS}}=\frac{8\pi^{(d-1)/2}}{(d+2)\Gamma(d/2)}n\sigma^{d-1}\sqrt{\frac{T}{m}}, \quad 
\nu_\eta^{*\text{IHS}}=\frac{(1+\al)\left[3\left(1-\al\right)+2d\right]}{2d(d+2)}, \quad 
\zeta^{*\text{IHS}}=\frac{1-\al^2}{2d}.
\end{equation}
%\beq
%\label{a7}
%\nu_\eta^*=\frac{3}{4d}(1+\al)\left(1-\al+\frac{2}{3}d\right), 
%\eeq
%\beq
%\label{a8}
%\zeta^{*\text{IHS}}=\frac{1-\al^2}{2d}.
%\eeq
The BGK-type kinetic model \eqref{4.5} yields the result
\beq
\label{a9}
\int d\mathbf{v}\; m V_i V_j  \Big[-\nu'\left(f-f^\text{M}\right)+\frac{\epsilon}{2}\frac{\partial}{\partial \mathbf{V}}\cdot \mathbf{V}f\Big]=-n T_g \Big[\left(\nu'+\epsilon\right)P_{ij}^*-\nu' \chi \delta_{ij}\Big],
\eeq
where we recall that $P_{ij}^*=P_{ij}/nT_g$. Comparison between Eqs.\ \eqref{a5} and \eqref{a9} yields Eq.\ \eqref{4.11} for $\nu'(\al)$.

%\begin{adjustwidth}{-\extralength}{0cm}

\end{document}